\documentclass[twocolumn,tighten]{aastex62}
\usepackage{amsmath,bm,graphicx,color} 

\newcommand{\vcz}{\mathrm{v}_z}
\newcommand{\vcy}{\mathrm{v}_y}
\newcommand{\vcx}{\mathrm{v}_x}

\newcommand{\vkb}{\boldsymbol{k}}
\newcommand{\vrb}{\boldsymbol{r}}
\newcommand{\bomega}{\boldsymbol{\Omega}}
\newcommand{\grad}{\boldsymbol{\nabla}}
\newcommand{\dvg}{\boldsymbol{\nabla}\boldsymbol{\cdot}}

\newcommand{\cnabla}{\!\boldsymbol{\cdot}\!\boldsymbol{\nabla}}

\newcommand{\bcdot}{\boldsymbol{\cdot}}
\newcommand{\bilinear}[2]{{#1}\!\boldsymbol{\cdot}\!{#2}}
\newcommand{\Ro}{\mathrm{Ro}}
\newcommand{\rmv}{v}

\newcommand{\Rowt}{\mathrm{\widetilde{Ro}_w}}
\newcommand{\Row}{\mathrm{Ro_w}}
\newcommand{\Roc}{\mathrm{Ro_c}}

\shorttitle{Rotating Convection: Penetration \& Gravito-inertial Waves}
\shortauthors{Augustson et al.}

\begin{document}

\title{A model of rotating convection in stellar and planetary interiors: II -  gravito-inertial wave generation} 

\correspondingauthor{K.~C. Augustson} 
\email{kyle.augustson@cea.fr}
\author{K.~C. Augustson}
\author{S. Mathis}
\author{A. Astoul}

\affil{AIM, CEA, CNRS, Universit\'{e} Paris-Saclay, Universit\'{e} Paris Diderot, Sarbonne Paris Cit\'{e}, F-91191 Gif-sur-Yvette Cedex, France}\label{inst1}

\begin{abstract}
  Gravito-inertial waves are excited at the interface of convective and radiative regions and by the Reynolds stresses
  in the bulk of the convection zones of rotating stars and planets. Such waves have notable asteroseismic signatures in
  the frequency spectra of rotating stars, particularly among rapidly rotating early-type stars, which provides a means
  of probing their internal structure and dynamics.  They can also transport angular momentum, chemical species, and
  energy from the excitation region to where they dissipate in radiative regions.  To estimate the excitation and
  convective parameter dependence of the amplitude of those waves, a monomodal model for stellar and planetary
  convection as described in Paper I is employed, which provides the magnitude of the rms convective velocity as a
  function of rotation rate. With this convection model, two channels for wave driving are considered: excitation at a
  boundary between convectively stable and unstable regions and excitation due to Reynolds-stresses. Parameter regimes
  are found where the sub-inertial waves may carry a significant energy flux, depending upon the convective Rossby
  number, the interface stiffness, and the wave frequency. The super-inertial waves can also be enhanced, but only for
  convective Rossby numbers near unity. Interfacially excited waves have a peak energy flux near the lower cutoff
  frequency when the convective Rossby number of the flows that excite them are below a critical Rossby number that
  depends upon the stiffness of the interface, whereas that flux decreases when the convective Rossby number is larger
  than this critical Rossby number.
\end{abstract}


\defcitealias{stevenson79}{S79}
\defcitealias{zahn91}{Z91}
\defcitealias{press81}{P81}

\section{Introduction}\label{sec:intro}

Gravito-inertial waves (hereafter GIWs) are low-frequency internal gravity waves (hereafter IGWs) that propagate in the
stably stratified regions of rotating stars and planets \citep{dintrans00}.  They propagate under the simultaneous
restoring action of the buoyancy and Coriolis forces.  Such waves are currently detected at the surface of rapidly
rotating intermediate-mass and massive stars thanks to high-precision asteroseismology \citep[e.g.,][and references
  therein]{neiner12b,moravveji16,vanreeth18,christophe18a,aerts18,aerts19}.  Moreover, GIWs and IGWs have been detected
through multiple observational techniques in the atmosphere, interior, and oceans of Earth
\citep[e.g.,][]{melchior86,gerkema08,gubenko18,maksimova18}, and the atmospheres of Mars \citep[e.g.,][]{gubenko15},
Jupiter \citep[e.g.,][]{young97,fletcher18}, Titan \citep[e.g.,][]{hinson83}, and Venus
\citep[e.g.,][]{tellmann12,ando18}. In intermediate-mass and massive stars, GIWs and IGWs constitute a powerful probe of
the chemical stratification and the radial differential rotation at the boundary between the convective core and the
radiative envelope \citep[e.g.,][]{vanreeth16,ouazzani17,vanreeth18,christophe18a,li19}.  While propagating in the
convectively stable zones of stars and planets, they are able to transport angular momentum, energy, and chemicals to
the regions where they dissipate through thermal diffusion \citep[e.g.,][]{schatzman93,zahn97,mathis08,mathis09},
co-rotation resonances \citep[e.g.,][]{goldreich90,alvan13}, and nonlinear wave breaking
\citep[e.g.,][]{rogers13,rogers17}. Thus, GIWs, alongside magnetic fields, provide a possible explanation for the weak
radial differential rotation revealed by space-based helioseismology and asteroseismology observations of stellar
radiative zones across the Hertzsprung-Russel diagram
\citep[e.g.,][]{garcia07,beck12,deheuvels12,mosser12,deheuvels14,kurtz14,benomar15,saio15,murphy16,spada16,vanreeth16,aerts17,fossat17,gehan18}.
Indeed, IGWs have been shown to be potentially efficient at angular momentum redistribution in the radiative core of the
Sun \citep[e.g.,][]{talon05,charbonnel13,mathis13a,mathis13b}, in sub-giant stars \citep{pincon17}, and in the radiative
envelope of early-type stars \citep[e.g.,][]{lee93,lee14,rogers15,fuller17,fuller18}.

Therefore, the excitation mechanisms and the resulting amplitudes and frequency spectrum need to be understood in an
astrophysical context.  In this work, the focus is on the stochastic excitation of GIWs at convective-radiative
interfaces and in the bulk of convective regions by turbulent Reynolds stresses.  Indeed, small-scale eddies or
large-scale turbulent structures such as convective plumes are able to perturb the interface between radiative and
convective zones, leading to the excitations of IGW and GIW packets.  The influence of small-scale eddies has been
semi-analytically been modelled for IGWs by \citet{press81} and \citet{zahn97} for example, whereas the impact of
larger-scale flows modelled analytically as collections of plumes on IGWs has been considered by \citet{schatzman93} and
\citet{pincon16} for example.  These excitation mechanisms have also been observed in 2D and 3D local and global
numerical simulations
\citep[e.g.,][]{hurlburt86,browning04,dintrans05,kiraga05,rogers05b,rogers06,rogers13,alvan14,alvan15,augustson16,edelmann19}. In
addition, turbulent Reynolds stresses in the bulk of convective regions also contribute to the generation of IGWs both
in late-type stars \citep[e.g.,][]{belkacem09a} and in early-type stars \citep[e.g.,][]{samadi10,shiode13} through their
coupling to the evanescent tail of the IGWs in the convective zone \citep[e.g.,][]{lecoanet13}.  This distributed effect
due to Reynolds stresses has been studied in the context of laboratory experiments of the temperature stratified
convective to non-convective transition of water as seen in \citet{lebars15,lecoanet15}, and \citet{couston18}, finding
that the Reynolds stresses are the dominant wave excitation mechanism in that system.

However, most of the above mentioned studies have neglected the action of rotation both on the turbulent convective
flows \citep[see e.g.,][; and references therein]{julien06,davidson13,brun17a,alexakis18} and on the IGWs that become
GIWs.  \citet{belkacem09b} have presented a formalism for the study of the stochastic excitation of IGWs in rotating
stars, although only in the case of slowly rotating stars.  Building upon this approach, \citet{mathis14} demonstrated
how the nature of the couplings between the GIWs and the turbulent Reynolds stresses could be strongly affected by the
Coriolis acceleration.  On one hand, those waves with frequencies above twice the rotation rate, super-inertial waves,
are evanescent in stellar convective regions, and thus only weakly couple to the Reynolds stresses away from the
convective-radiative transition.  On the other hand, those waves with frequencies below twice the rotation rate,
sub-inertial waves, become propagative inertial waves in stellar convection zones and are intrinsically coupled with the
turbulent convective flows throughout the convection zone. The reader is referred to the detailed discussion of this in
\citet{mathis14}. Moreover, turbulent structures become strongly anisotropic with global alignment with the rotation
axis while the efficiency of the heat transfer between different scales is globally decreased
\citep[e.g.,][]{sen12,julien12}.  Additionally, turbulent convective structures can be understood as being a combination
of inertial waves in the asymptotic regime of rapid rotation \citep[e.g.,][]{davidson13,clarkdileoni14}.  These
mechanisms can be very important in stars since late-type stars are rapidly rotating during their pre-main-sequence
phase \citep[e.g.,][]{gallet15}, while early-type stars generally have high rotation rates throughout their evolution
\citep[e.g.,][; and references therein]{maeder00}.  Yet, \citet{mathis14} did not provide a quantitative estimate of the
GIW amplitudes, frequency spectrum, and induced transport of momentum and chemicals due to the lack of a prescription
for rotating turbulent convection.

\citet{stevenson79} and \citet{augustson19b} (hereafter Paper I) have derived mixing-length based scaling laws for the
primary properties of small-scale convective eddies in rotating stellar and planetary convection (e.g., their rms
velocity, horizontal convective scale, and the local superadiabaticity).  Direct nonlinear f-plane numerical simulations
of \citet{kapyla05} and \citet{barker14} have shown that these prescriptions appear to hold up well in polar regions.
Thus, the convective scaling laws are employed to provide a first quantitative analytical estimate of the amplitudes and
frequencies of stochastically excited GIWs.  Indeed, such a model permits the action of (rapid) rotation to be taken
into account both for the propagation of GIWs and for the nature of the convection.  The obtained formalism constitutes
a generalization of the work of \citet{lecoanet13} for pure IGWs in the nonrotating case.  This formalism can be
implemented into stellar evolution and oscillation codes to explore the properties and consequences of GIWs across the
Hertzsprung-Russell diagram.  Therefore, this represents a step toward building a coherent theoretical framework to
study the seismology of rotating stars and the wave-induced transport in their interiors, working synergistically with
the ongoing development of numerical simulations and laboratory experiments. For instance, see the recent laboratory
experiments by \citet{rodda18}.

\subsection{Outline}

The model of convection derived in Paper I is employed to estimate the GIW energy flux into the stable region adjacent
to convective zones. The general framework of the convection model is briefly summarized in \S\ref{sec:genframe}. GIWs
and their excitation mechanisms are briefly reviewed in \S\ref{sec:introwaves}. Following the arguments of
\citet{press81} and \citet{andre17}, the interfacial generation of GIWs and their associated energy flux is assessed in
\S\ref{sec:interfacewaves}. Subsequently, in \S\ref{sec:reynoldswaves}, an estimate is given for the energy flux of GIWs
excited by Reynolds stresses using the convection model. A summary of the results and perspectives are presented in
\S\ref{sec:final}.

\section{Heat-Flux Maximized Convection Model}\label{sec:genframe}

\subsection{Hypotheses and Localization}

A self-consistent and yet computationally tractable treatment of stellar and planetary convection has been a long sought
goal, with many such models having been employed in evolution models.  One such model based upon a variational principle
for the maximization of the heat flux \citep{howard63} and a turbulent closure assumption for the velocity amplitude
\citep{stevenson79} has been expanded upon in Paper I \citep{augustson19b}.  In the context of GIW excitation, one needs
to ascertain the amplitude of the velocity field that excites the waves both through Reynolds stresses acting throughout
the bulk of the convection zone on both the evanescent super-inertial waves and propagating inertial waves and also
exciting them directly through thermal buoyancy in the region of convective penetration.

\begin{figure}[t!]
  \begin{center}
    \includegraphics[width=0.5\textwidth]{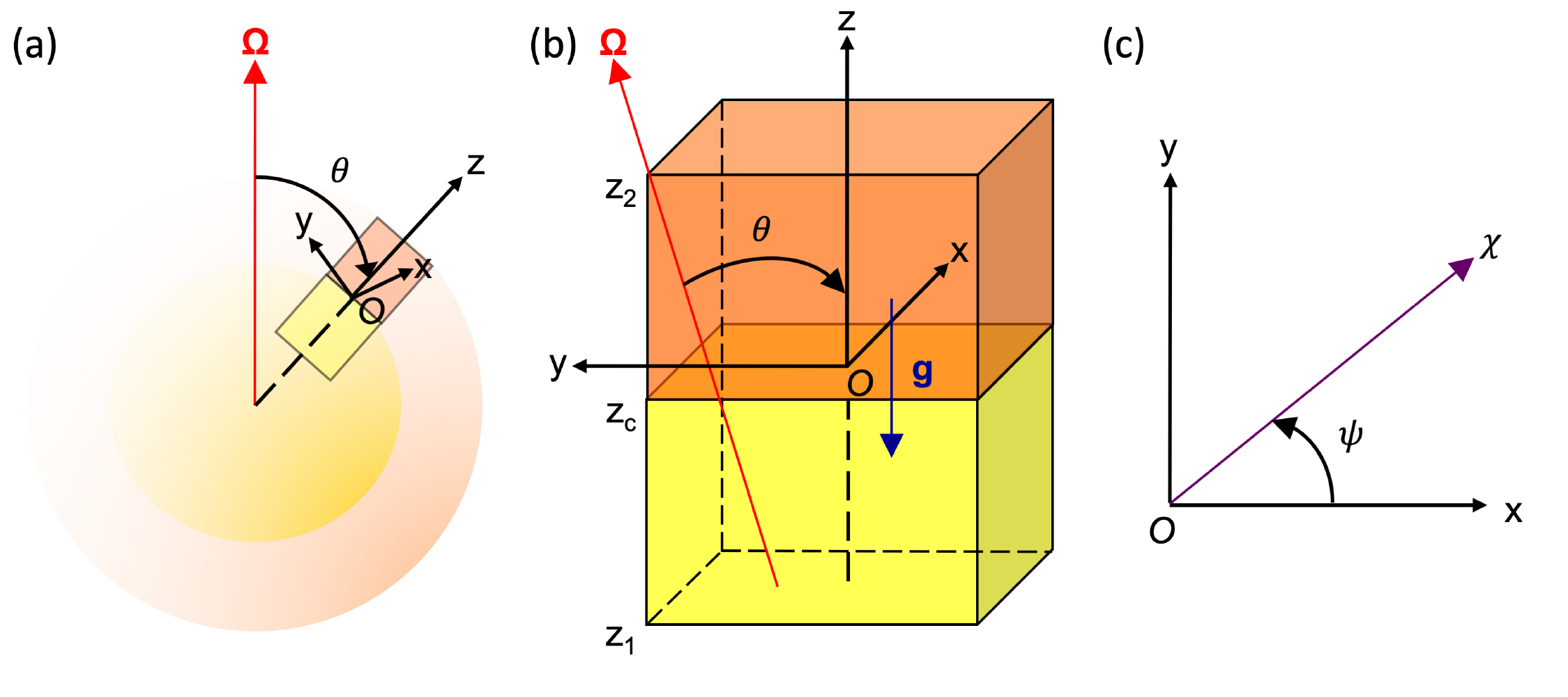} 
    \caption{Coordinate system adopted for the models of rotating convection and gravity-wave excitation, showing (a)
      the global geometry and f-plane localization, (b) the f-plane geometry, and (c) the direction $\chi$ in the
      horizontal plane of the f-plane.  The orange tones denote a convective region and the yellow tones denote a stable
      region for late-type stars, and vice versa in early-type stars. }\label{fig:coords}
  \end{center}
\end{figure}

To that end, a local region is considered as in Paper I, where a small 3D section of the spherical geometry is the focus
of the analysis.  This region covers a portion of both the convectively stable and unstable zones as shown in Figure
\ref{fig:coords}, where the set up is configured for a low mass star with an external convective envelope. One may
exchange these regions when considering a more massive star with a convective core. In this local frame, there is an
angle between the effective gravity $\mathbf{g}_{\mathrm{eff}}$ and the local rotation vector that is equivalent to the
colatitude $\theta$.  The Cartesian coordinates are defined such that the vertical direction $z$ is anti-aligned with
the gravity vector, the horizontal direction $y$ lies in the meridional plane and points toward the north pole defined
by the rotation vector, the horizontal direction $x$ is equivalent to the azimuthal direction.  The angle $\psi$ in the
horizontal plane defines the direction of horizontal wave propagation $\chi$.

While the details of the derivation of the heat-flux maximized rotating convection model may be found in Paper I, it is
necessary to recall a few of the relevant results as they are applied in subsequent sections.  The heuristic model is
local such that the length scales of the flow are much smaller than either the density or pressure scale heights, thus
ignoring the global-scale flows, which will be the focus of a forthcoming paper.  The dynamics are further considered to
be in the Boussinesq limit.  This localization of the convection therefore consists of an infinite layer of a nearly
incompressible fluid with a small thermal expansion coefficient $\alpha_T=-\partial\ln{\rho}/\partial T|_P$ that is
confined between two infinite impenetrable boundaries differing in temperature by $\Delta T=T(z_2)-T(z_c)$, with the
lower boundary located at $z_1$ and the upper boundary at $z_2$.  In this model, it is assumed for this model that
$T(z_2)<T(z_c)$ and that the boundaries are separated by a distance $\ell_0=z_2-z_c$, as in Figure \ref{fig:coords},
where $z_c$ is the point of transition between the convectively stable and unstable regions.

The recent motivation behind the development of the convection model arose from the numerical work of \citet{kapyla05}
and \citet{barker14}, where it was found that the rotational scaling of the amplitude of the temperature, its gradient,
and the velocity field compare well with those derived in \citet{stevenson79}.  Moreover, the experimental work of
\citet{townsend62} and the analysis of \citet{howard63} have shown that a heat-flux maximization principle
provides a sound basis for the description of Rayleigh-B\'{e}nard convection, leading to its use here. Thus, two
hypotheses underlie the convection model: the Malkus conjecture that the convection arranges itself to maximize the heat
flux and that the nonlinear velocity field can be characterized by the dispersion relationship of the linearized
dynamics. Constructing the model of rotating convection then consists of three steps: deriving a dispersion relationship
that links the normalized growth rate $\hat{s}=s/N_*$ to $q = N_{*,0}/N_*$, which is the ratio of superadiabaticity of
the nonrotating case to that of the rotating case (where $N_*^2=|g\alpha_T\beta|$ is the absolute value of the square of
the Brunt-V\"{a}is\"{a}l\"{a} frequency), and to the normalized wavevector $\xi^3=k^2/k_z^2$, maximizing the heat flux
with respect to $\xi$, and assuming an invariant maximum heat flux that then closes this three variable system.

\subsection{Dispersion Relationship and Flux Maximization}

For rotating convection, one may show that for impenetrable and stress-free boundary conditions the solutions of the
equations of motion are periodic in the horizontal, sinusoidal in the vertical, and exponential in time, e.g.
$\vcz = \rmv \sin{\left[k_z\left(z-z_c\right)\right]}\exp{(i\vkb_\perp\bcdot\vrb + s t)}$, where $\vkb_\perp$ is the
horizontal wavevector, $s$ is the growth rate, $\vrb$ is the local coordinate vector, and $\rmv$ is a constant velocity
amplitude. To satisfy the impenetrable, stress-free, and fixed temperature boundary conditions, it is required that the
vertical wavenumber be $k_z=n\pi/\ell_0$. The introduction of this solution into the reduced linearized equation of
motion yields the following dispersion relationship that relates $s$ to the wavevector $\vkb$ as

\vspace{-0.25truein}
\begin{center}
  \begin{align}
    &\left(s +\kappa k^2\right)\left(s + \nu k^2\right)^2 k^2 + 
            g\alpha_T\beta k_{\perp}^2\left(s + \nu k^2\right) \nonumber \\
    &\qquad+ 4\left(\bilinear{\bomega}{\vkb}\right)^2\left(s + \kappa k^2\right) = 0. \label{eqn:fvcz}
  \end{align}
\end{center}

\noindent This equation may be nondimensionalized by dividing through by the appropriate powers of $N_*$ and $k_z$,
leading to the definition of additional quantities

\vspace{-0.25truein}
\begin{center}
  \begin{align}
    \hat{s} &=\frac{s}{N_*}, \,\,\,
    \xi^3 = 1+ a^2 = \frac{k^2}{k_z^2}, \,\,\,
    a^2 = \frac{k_x^2}{k_z^2}+\frac{k_y^2}{k_z^2} = a_x^2 + a_y^2, \label{eqn:defszakv}\\
    K &= \frac{\kappa k_z^2}{N_*}, \quad
    V = \frac{\nu k_z^2}{N_*}. \nonumber
  \end{align}
\end{center}

\noindent Introducing these into the dispersion relationship yields

\vspace{-0.25truein}
\begin{center}
  \begin{align}
    &\left(\hat{s} \!+\! K z^3\right)\!\! \left(\! \xi^3\!\left(\hat{s}\!+\! V \xi^3\right)^2 \!\!+\! O^2 \left(\cos{\theta}
      + a_y\sin{\theta}\right)^2\!\right) \nonumber\\
    &-\! \left(\xi^3\!-\! 1\right)\!\!\left(\hat{s}\!+\! V \xi^3\right)\!=\!0, \label{eqn:fullchar}
  \end{align}
\end{center}

\noindent with $4(\bomega\cdot\vkb)^2/N_*^2= k_z^2 O^2 \left(\cos{\theta} + a_y\sin{\theta}\right)^2$ where

\vspace{-0.25truein}
\begin{center}
  \begin{align}
    O^2 &= \frac{4\Omega_0^2}{N_*^2},
  \end{align}
\end{center}

\noindent where $\Omega_0$ is the bulk rotation rate of the system.

The characteristic velocity $\rmv_0$ of the nonrotating and nondiffusive case is derived from the growth rate and
maximizing wavevector in that case, with $s_0^2=3/5|g_0\alpha_T\beta_0|=\left(3/5\right)N^2_{*,0}$, $\beta_0$ being the
thermal gradient, $g_0$ being the effective gravity, and where $k_0^2=\left(5/2\right) \,k_z^2$ with $k_z = \pi/\ell_0$.  This leads to

\vspace{-0.25truein}
\begin{center}
  \begin{align}
    \rmv_0 = \frac{s_0}{k_0} = \frac{\sqrt{6}}{5} \frac{N_{*,0}}{k_z} =  \frac{\sqrt{6}}{5\pi} N_{*,0}\ell_0.\label{eqn:v0}
  \end{align}
\end{center}

\noindent Thus, the definition of the convective Rossby number $\Roc$ is

\vspace{-0.25truein}
\begin{center}
  \begin{align}
    \Roc=\frac{\rmv_0}{2\Omega_0 \ell_0}=\frac{\sqrt{6} N_{*,0}}{10\pi\Omega_0}, \label{eqn:rossby}
  \end{align}
\end{center}

\noindent which implies that

\vspace{-0.25truein}
\begin{center}
  \begin{align}
    O &=\frac{2\Omega_0}{N_*} = \frac{\rmv_0}{N_* \Roc\ell_0} = \frac{\sqrt{6}N_{*,0}}{5\pi N_*\Roc}. \label{eqn:o0def}
  \end{align}
\end{center}

The superadiabaticity for this system is $\epsilon = H_P \beta/T$, meaning that $N_*^2 = |g \alpha_T T \epsilon/H_P|$,
where $H_P$ is the pressure scale height. The potential temperature gradient in the nonrotating and nondiffusive case is
ascertained from the Malkus-Howard turbulence model \citep{malkus54,howard63}, which yields a value of $N_{*,0}$. It is
also useful to compare the timescales relative to $N_{*,0}$.  Letting the ratio of superadiabaticities be

\vspace{-0.25truein}
\begin{center}
  \begin{align}
	q &= N_{*,0}/N_*,\label{eqn:qdef}
  \end{align}
\end{center}

\noindent all parametric quantities have the following equivalencies

\vspace{-0.25truein}
\begin{center}
  \begin{align}
    O &= q \frac{\sqrt{6}}{5\pi\Roc}= q O_0,\nonumber\\
    K &= q \frac{\kappa k_z^2}{N_{*,0}} = q K_0,\label{eqn:equivalencies}\\
    V &= q \frac{\nu k_z^2}{N_{*,0}} = q V_0.\nonumber
  \end{align}
\end{center}

So, the dispersion relationship (Equation \ref{eqn:fullchar}) and the heat flux may be
written as

\vspace{-0.25truein}
\begin{center}
  \begin{align}
    &\left(\hat{s} \!+\! K_0 q \xi^3\right)\!\! \left(\! \xi^3\!\left(\hat{s}\!+\! V_0 q \xi^3\right)^2 \!\!+\! O_0^2q^2\cos^2{\theta}\!\right) \!\nonumber\\
    &\qquad-\! \left(\xi^3\!-\! 1\right)\!\!\left(\hat{s}\!+\! V_0 q \xi^3\right)\!=\!0, \label{eqn:fullcharq}\\
    &F = \frac{F_0}{q^3} \left[\frac{\hat{s}^3}{\xi^3}+V_0 q\hat{s}^2\right],\label{eqn:heatfluxq}
  \end{align}
\end{center}

\noindent where $F_0 = \langle\rho\rangle c_P N_{*,0}^3/\left(g\alpha_T k_z^2\right)$.

To ascertain the scaling of the superadiabaticity, the velocity, and the horizontal wavevector with rotation and
diffusion, an additional assumption is made to close the system.  This assumption is that the maximum heat flux is invariant
to any parameters: $\max{\left[F\right]}=\max{\left[F\right]}_0$ so the heat flux is equal to the maximum value
$\max{\left[F\right]}_0$ obtained for the nonrotating case, which fits with the assumption that the energy generation of
the star is not strongly effected by rotation.

In the case of planetary and stellar interiors, the viscous damping timescale is generally longer than the convective
overturning timescale (e.g., $V_0\ll N_{*,0}$).  Thus, the maximized heat flux invariance is much simpler to treat.  In
particular, the heat flux invariance condition under this assumption is then

\vspace{-0.25truein}
\begin{center}
  \begin{align}
    \frac{\max{\left[F\right]}}{\max{\left[F\right]}_0}
    &=\frac{25}{6}\sqrt{\frac{5}{3}}\left[\frac{\hat{s}^3}{q^3\xi^3}+\frac{V_0\hat{s}^2}{q^2}\right]_{\mathrm{max}}\nonumber\\
    &\approx\left.\frac{25}{6}\sqrt{\frac{5}{3}}\frac{\hat{s}^3}{q^3\xi^3}\right|_{\mathrm{max}} =1,
  \end{align}
\end{center}

\noindent implying that

\vspace{-0.25truein}
\begin{center}
  \begin{align}    
    &\hat{s}=\tilde{s} q \xi + \mathcal{O}(V_0),\label{eqn:maxndhf}
  \end{align}
\end{center}

\noindent where $\tilde{s}=2^{1/3} 3^{1/2} 5^{-5/6}$ and $\max{\left[F\right]}_0=6/25\sqrt{3/5} F_0$ follows from the
definition of the flux and the maximizing wavevector used to define $\rmv_0$ above in Equation \ref{eqn:v0}.

\subsection{Rotational Scaling of Superadiabaticity, Velocity, and Wavevector}

The assumption of this convection model is that the magnitude of the velocity is defined as the ratio of the maximizing
growth rate and wavevector. With the above approximation, the velocity amplitude can be defined relative to the
nondiffusive and nonrotating case scales without a loss of generality as

\vspace{-0.25truein}
\begin{center}
  \begin{align}
    \frac{\rmv}{\rmv_0} &= \frac{k_0}{s_0}\frac{s}{k} = \frac{5}{\sqrt{6}} \frac{N_*}{N_{*,0}}\frac{\hat{s}}{\xi^{3/2}} =
                          \frac{5}{\sqrt{6}} \frac{\hat{s}}{q \xi^{3/2}} = \left(\frac{5}{2}\right)^{\frac{1}{6}} \xi^{-\frac{1}{2}}.\label{eqn:vsteve}
  \end{align}
\end{center}

\noindent So only the maximizing wavevector needs to be found in order to ascertain the relative velocity amplitude. For
reference, the symbols that will be frequently used from this section are listed in Table \ref{tab1}.

\begin{table}
  \begin{center}
    \begin{tabular}{Ll}
      \hline
      \hline
      $a=k_x/k_z$ & Maximizing horizontal wavevector\\
      $k_z=\pi/\ell_0$ & Maximizing vertical wavevector \\
      $K_0=\kappa k_z^2/N_{*,0}$ & Normalized thermal diffusivity\\
      $O_0=\sqrt{6}/\left(5\pi\Roc\right)$ & Normalized Coriolis coefficient\\
      $\Roc=\sqrt{6}N_{*,0}/\left(10\pi\Omega_0\right)$ & Convective Rossby number\\
      $\hat{s}=s/N_*$ & Normalized growth rate\\
      $\rmv_0=\sqrt{6}N_{*,0}\ell_0/\left(5\pi\right)$ & Velocity of the nonrotating case\\
      $V_0= \nu k_z^2/N_{*,0}$ & Normalized viscosity\\
      $q=N_{*,0}/N_*$ & Ratio of buoyancy timescales\\
      $\xi^3 = k^2/k_z^2$ & Normalized wavevector\\
      \hline
    \end{tabular}
    \caption{Frequently used symbols in the convection model.}\label{tab1}
  \end{center}
\end{table}

With all the equations in hand, the horizontal wavevector may be seen to be the
roots of the fourteenth-order polynomial,

\vspace{-0.25truein}
\begin{center}
  \begin{align}
    &\xi^3\! \left(V_0 \xi^2\!+\! \tilde{s}\right)^2\! \left[3V_0 K_0 \xi^4\!\left(2 \xi^3\!-\! 3\right)\right.\nonumber\\
    &\qquad\qquad\left.+\tilde{s} \xi^2\!\left(V_0\!+\! K_0\right)\!\left(4\xi^3\!-\! 7\right)\!+\! \tilde{s}^2\left(2 \xi^3\!-\! 5\right)\right]\nonumber\\
    &-\frac{6 \cos^2{\!\theta}}{25 \pi^2 \Ro_{\mathrm{c}}^2}\!\left[2\tilde{s}\left(K_0-V_0\right)+3\tilde{s}^2\xi\right.\nonumber\\
    &\qquad\qquad\left.+\tilde{s}\left(K_0+5V_0\right)\xi^3+3K_0V_0\xi^5\right]\!=\!0,\label{eqn:zeqndiff}
  \end{align}
\end{center}

whereas the superadiabaticity is defined as

\vspace{-0.25truein}
\begin{center}
  \begin{align}
    &\frac{\epsilon}{\epsilon_0} \!=\! \frac{\left(\tilde{s}\!+\!K_0 \xi^2\right)\! \left(25 \pi^2 \Ro_{\mathrm{c}}^2
      \tilde{s}^2\xi^5\!\left(\tilde{s}\!+\! V_0 \xi^2\right)^2\!+\! 6 \cos^2{\theta}\right)}{25\pi^2 \Ro_{\mathrm{c}}^2 \tilde{s}\left(\xi^3\!-\!
      1\right)\left(\tilde{s}\!+\! V_0 \xi^2\right)}. \label{eqn:xeqnfd}
  \end{align}
\end{center}

For the study of adiabatic GIWs, the nondiffusive model is employed where $V_0\rightarrow0$ and $K_0\rightarrow0$,
leading to

\vspace{-0.25truein}
\begin{center}
  \begin{align}
    2\xi^5-5\xi^2-\frac{18\cos^2{\theta}}{25\pi^2\Ro_{\mathrm{c}}^2\tilde{s}^2}=0,
  \end{align}
\end{center}

and

\vspace{-0.25truein}
\begin{center}
  \begin{align}
    &\frac{\epsilon}{\epsilon_0} = \frac{25\pi^2\Ro_{\mathrm{c}}^2\tilde{s}^2\xi^5+6\cos^2{\theta}}{25\pi^2\Ro_{\mathrm{c}}^2\tilde{s}^2\left(\xi^3-1\right)}. 
  \end{align}
\end{center}

So, to ascertain the maximizing wavenumber, and thus the velocity and superadiabaticity, of the motions that maximize the
heat flux one supplies the colatitude $\theta$ and the convective Rossby number of the flow $\Ro_{\mathrm{c}}$.  Now
that the quantities related to the convection model have been defined, the impact of rotation on the convective
excitation of gravito-inertial waves can be characterized.

\section{Gravito-Inertial Waves}\label{sec:introwaves}

When examining the excitation of GIWs, the region of interest is near the radiative-convective interface. As a first
step toward a coherent global treatment of GIW excitation, the forthcoming analysis will share the same Cartesian
geometry as the convection model, which is depicted in Figure \ref{fig:coords} where the stable region is now also
considered.  For compactness, one may introduce the two components of the rotation vector along the vertical direction
$z$ and the latitudinal direction $y$ as

\vspace{-0.25truein}
\begin{center}
  \begin{align}
    f=2\Omega_0 \cos{\theta}, \quad \text{and} \quad f_s=2\Omega_0 \sin{\theta}.
  \end{align}
\end{center}

\noindent As depicted in Figure \ref{fig:coords}, the waves to be considered propagate along a direction with an angle
$\psi$ in the horizontal $x-y$ plane, the latitudinal component of the rotation vector has two images in this plane with

\vspace{-0.25truein}
\begin{center}
  \begin{align}
    f_{sc}=2\Omega_0 \sin{\theta}\cos{\psi}, \quad \text{and} \quad f_{ss}=2\Omega_0 \sin{\theta}\sin{\psi}.\label{eqn:fss}
  \end{align}
\end{center}

In this analysis, both components of the rotation vector are kept in the equations of motion, as opposed to the
so-called traditional approximation that considers only its vertical component in the Coriolis acceleration in order to
yield a separable dynamical system. However, in the near-inertial frequency range, nontraditional effects act as a
singular perturbation.  Specifically, the phase of the wave has a vertical dependence that is absent under the
traditional approximation.  Also, as shown in \citet{gerkema05}, when considering a non-constant stratification,
sub-inertial GIWs can be trapped in regions of weak stratification. This behavior does not arise in the traditional
approximation.

The near-inertial wave dynamics are quite sensitive to variations in the effective Coriolis parameters $f$ and $f_s$,
which could arise from a locally strong vortex.  For instance, the low Rossby number, quasi-geostrophic flows that
likely exist deep in stellar interiors and that impinge upon stable regions could transform a near-inertial wave from
the super-inertial regime into the sub-inertial regime. The wave would suddenly find itself trapped in a waveguide,
leading to a strong interaction between the near-inertial waves and large-scale motions.  Such notions will be
considered in a forthcoming investigation of global-scale dynamics.

Following \citet{gerkema05} and \citet{mathis14}, the linearized equations of motion used to construct the convection
model above are extended into the radiative region to study the coupling of the convection with both the gravity and
inertial waves present in both regions.  Specifically, these equations are Boussinesq and in the Emden-Cowling
approximation \citep{emden07,cowling41}, where the gravitational potential perturbations are ignored, with

\vspace{-0.25truein}
\begin{center}
  \begin{align}
    \partial_t\vcx - f \vcy+f_s \vcz = - \partial_x p,\\
    \partial_t\vcy + f \vcx = - \partial_y p,\\
    \partial_t\vcz + f \vcx = - \partial_z p + b,\\
    \partial_x \vcx + \partial_y \vcy + \partial_z \vcz = 0,\\
    \partial_t b + N_R^2\left(z\right)\vcz  =0,
  \end{align}
\end{center}

\noindent where the buoyancy is $b=-g_{\mathrm{eff}}\rho'\left(\vrb,t\right)/\rho_0$. One may eliminate
the pressure, buoyancy and the horizontal velocities to yield an equation of motion for the vertical component of the
velocity as

\vspace{-0.25truein}
\begin{center}
  \begin{align}
    \left[\partial_t^2\nabla^2 + 4\left(\bomega\cnabla\right)^2+N_R^2\nabla_\perp^2\right]\vcz=0,
  \end{align}
\end{center}

\noindent where $\nabla_\perp^2$ is the horizontal Laplacian and $N_R^2\left(z\right)$ is the Brunt-V\"{a}is\"{a}l\"{a}
frequency in the radiative zone.  If one then further considers monochromatic GIWs with a frequency $\omega$ that
propagates along the direction characterized by the angle $\psi$ in the horizontal plane and a coordinate
$\chi = x\cos{\psi} + y\sin{\psi}$ along that direction as in Figure \ref{fig:coords} with a solution of the form
$v_z\left(\vrb,t\right) = w\left(\vrb\right) e^{i\omega t}$, one obtains the Poincar\'{e} equation for the GIWs

\vspace{-0.25truein}
\begin{center}
  \begin{align}
    \left(N_R^2-\omega^2+f_{ss}^2\right)\partial_\chi^2w + 2f f_{ss} \partial_{\chi z}w + \left(f^2-\omega^2\right)\partial_z^2w=0.
  \end{align}
\end{center}

\noindent Nominally, this is a nonseparable equation.  However, it may be transformed when assuming the following
spatial form of the solution

\vspace{-0.25truein}
\begin{center}
  \begin{align}
    w = \widehat{w}\left(z\right)e^{ik_\perp \left[\chi + \delta\left(\Row\right) z\right]}, \label{eqn:form}
  \end{align}
\end{center}

\noindent as in \citet{gerkema05}, where $k_\perp$ is the wavevector along $\chi$, $\Row=\omega/2\Omega_0$ is the wave
Rossby number, and

\vspace{-0.25truein}
\begin{center}
  \begin{align}
    \displaystyle{\delta(\Row) =
      \frac{\sin{\!\theta}\cos{\!\theta}\sin{\!\psi}}{\Ro_{\mathrm{w}}^{2}-\cos^2{\!\theta}}}
  \end{align}
\end{center}

\noindent is the phase shift linking the horizontal and vertical directions.  Yet the above form of the solution leads
to a homogeneous Schr\"{o}dinger-like equation in the vertical coordinate as

\vspace{-0.25truein}
\begin{center}
  \begin{align}
    \partial_z^2\widehat{w} + k_V^2\left(z\right) \widehat{w} = 0,\label{eqn:poinred}
  \end{align}
\end{center}

\noindent where

\vspace{-0.25truein}
\begin{center}
  \begin{align}
    k_V^2\left(z\right) = k_\perp^2 \left[\frac{N_R^2-\omega^2}{\omega^2-f^2}+\left(\frac{\omega f_{ss}}{\omega^2-f^2}\right)^2\right].
  \end{align}
\end{center}

\noindent Similar to the nonrotating case, this permits the use of the method of vertical modes to find the modal
functions $\widehat{w}_j$ that satisfy the appropriate boundary conditions. Indeed, it can be shown that solutions of
the form of Equation \ref{eqn:form} constitute an orthogonal and complete basis \citep{gerkema05}.

In convectively stable regions where rotation is important, GIWs may propagate if their frequency falls within the range
between $\omega_{-}$ and $\omega_{+}$,

\vspace{-0.25truein}
\begin{center}
  \begin{align}
	\omega_\pm \!=\! \frac{1}{\sqrt{2}}\sqrt{N_R^2\! +\! f^2\! +\! f_{ss}^2 \pm \!\sqrt{\left(N_R^2\! +\! f^2 \!+\! f_{ss}^2\right)^2 - \left(2N_R f\right)^2}}, \label{eqn:giwdisper}
  \end{align}
\end{center}

\noindent whereas in convection zones one has that

\vspace{-0.25truein}
\begin{center}
  \begin{align}
    k_{\mathrm{CZ}}^2 = k_\perp^2 \frac{\widetilde{\Row}^{-2}-1}{\left(1-\Row\cos^2{\theta}\right)^2},
  \end{align}
\end{center}

\noindent where the local wave Rossby number is

\vspace{-0.25truein}
\begin{center}
  \begin{align}
    \displaystyle{\Rowt\!=\!\frac{\Row}{\sqrt{\cos^2{\!\theta}\!+\!\sin^2{\!\theta}\sin^2{\!\psi}}}}.
  \end{align}
\end{center}

At the pole in a convectively stable region, this implies that the frequency must be between $2\Omega_0$ and $N_R$ for
the wave to propagate, where the Brunt-V\"{a}is\"{a}l\"{a} frequency is typically much larger than the rotational
frequency in the radiative core of late-type stars and the radiative envelope of early-type stars
\citep[e.g.,][]{aerts10}. More generally, at other latitudes, the hierarchy of extremal propagative wave frequencies
satisfy the inequality $\omega_{-}<2\Omega_0<N_R<\omega_{+}$.  As these waves propagate, the Brunt-V\"{a}is\"{a}l\"{a}
frequency varies, for instance it becomes effectively zero in the convection zone. This implies that waves in the
frequency range $\omega<2\Omega_0$ are classified as sub-inertial GIWs in stable regions, becoming pure inertial waves
in convective regions. Waves in the frequency range $\omega\ge 2\Omega_0$ are classified as super-inertial GIWs in the
stable region, which in contrast to sub-inertial GIWs become evanescent in the convective region.  Figure 2 in
\citet{mathis14} provides a concise visual reference of the hierarchy of frequencies, to which the reader is referred.

\section{Interfacial Gravito-Inertial Wave Energy Flux Estimates}\label{sec:interfacewaves}

There are many models for estimating the magnitude of the gravity wave energy flux arising from the waves excited by
convective flows. One of the first and most straightforward of such estimates is described in \citet[][hereafter
P81]{press81}, where the wave energy flux across an interface connecting a convective region to a stable zone is
computed by matching their respective pressure perturbations at that interface. Because the wave excitation occurs at an
interface, the pressure perturbations are more important than the Reynolds stresses of the flows. What is more, the
model assumes that the convective source is a delta function in Fourier space. So, the model permits only a single
horizontal spatial scale $2\pi/k_c$ and a single time scale for the convection $2\pi/\omega_c$ that also selects the
depth of the transitional interface where $N_R(r)=\omega_c$ for gravity waves, where $\omega_c = \omega_0 /\sqrt{\xi}$
with $\omega_0 = 2\pi \rmv_0/\ell_0$, which lends itself well to the above convection model. This approach yields a wave
energy flux proportional to the product of the convective kinetic energy flux and the ratio of the wave frequency to the
Brunt-V\"{a}is\"{a}l\"{a} frequency in the nonrotating case for gravity waves.

The convective model established above captures some aspects of the influence of rotation on the convective
flows. Therefore, the impact of the Coriolis force on the stochastic excitation of GIWs can be evaluated. In this
context, recent work has established an estimate of the GIW energy flux \citep{andre17}.  It can be used to estimate the
rotational scaling of the amplitude of the wave energy flux arising from the modified properties of the convective driving.  From
Equation 61 of \citet{andre17}, the vertical GIW energy flux can be computed from the horizontal average of the product of the vertical
velocity and pressure perturbation that, given the linearization of the Boussinesq equations for monochromatic waves
propagating in a selected horizontal direction, can be evaluated to be

\vspace{-0.25truein}
\begin{center}
  \begin{align}
    F_z=\frac{1}{2}\rho_0 \frac{\omega^2-f^2}{\omega k_\perp^2} k_z \rmv_w^2,
  \end{align}
\end{center}

\noindent where $\rmv_w$ is the magnitude of the vertical velocity of the wave.  Moreover, the solution for the vertical
velocity implies that the dispersion relationship is

\vspace{-0.25truein}
\begin{center}
  \begin{align}
    \frac{k_z}{k_\perp} = \left[\frac{N_R^2-\omega^2}{\omega^2-f^2}+\left(\frac{\omega f_{ss}}{\omega^2-f^2}\right)^2\right]^{\frac{1}{2}},
  \end{align}
\end{center}

\noindent where $f$ and $f_{ss}$ are defined above in Equation \ref{eqn:fss}. Note that a reference table is
given to help identify the many parameters in this section (Table \ref{tab2}).

\begin{table}
  \begin{center}
    \begin{tabular}{Ll}
      \hline
      \hline
      $\chi=\cos{\psi}x+\sin{\psi}y$ & Horizontal position\\
      $\displaystyle{\delta(\Row) =
      \frac{\sin{\!\theta}\cos{\!\theta}\sin{\!\psi}}{\Ro_{\mathrm{w}}^{2}-\cos^2{\!\theta}}}$ & Horizontal phase shift\\
      $\Delta_j=\int_{z_j}^{z_c}dk_\perp |k_V|$ & Vertical phase shift\\
      $k_\perp\!=\cos{\psi}k_x+\sin{\psi}k_y$ & Horizontal wavevector\\
      $k_V$ & Vertical wavevector\\
      $f=2\Omega_0\cos{\theta}$ & V. Coriolis frequency\\
      $f_s=2\Omega_0\sin{\theta}$ & H. Coriolis frequency\\
      $f_{ss}=2\Omega_0\sin{\theta}\sin{\psi}$ & P.~H. Coriolis frequency\\
      $f_{sc}=2\Omega_0\sin{\theta}\cos{\psi}$ & P.~H. Coriolis frequency\\
      $N_R$ & Brunt-V\"{a}is\"{a}l\"{a} frequency\\
      $\omega_c=\sqrt{6}v N_{*,0} /\left(5\pi v_0\right)$ & Convective frequency\\
      $\omega_e$ & Eddy timescale\\
      $\psi$ & Angle in horizontal plane\\
      $\Row=\omega/2\Omega_0=5\pi\sigma \Roc S/\sqrt{6}$ & Wave Rossby number\\
      $\displaystyle{\Rowt\!=\!\frac{\Row}{\sqrt{\cos^2{\!\theta}\!+\!\sin^2{\!\theta}\sin^2{\!\psi}}}}$ & Local Rossby number\\
      $S=N_R/N_0$ & Interface stiffness\\
      $\mathbb{S}$ & Convective source\\
      $\sigma=\omega/N_R$ & Normalized frequency\\
      \hline
    \end{tabular}
    \caption{Frequently used symbols in the models of interfacially-excited and Reynolds-stress-induced wave
      excitation. The abbreviations used in the table are V. for Vertical, H. for Horizontal, and P.~H. for
      Projected Horizontal.}\label{tab2}
  \end{center}
\end{table}

Following \citetalias{press81}, further assumptions are necessary to complete the estimate of the wave energy flux.  The
convection is turbulent. So the fluctuating part of the velocity field is of the same order of magnitude as the
convective eddy turnover velocity $\rmv \approx \omega_c/k_c$, which implies that convective pressure perturbations
are approximately $P_c = \rho_0 \rmv^2$. Assuming that the pressure is continuous across the interface between the
convectively stable and unstable regions, the horizontally-averaged pressure perturbations of the propagating waves
excited at the interface must then be equal to the turbulent pressure on the convective side of the interface. Those
pressure perturbations follow from the solution for the vertical velocity and the nondiffusive Boussinesq equations
\citep{andre17}. For plane wave solutions, the magnitude of those perturbations can then be written as

\vspace{-0.25truein}
\begin{center}
  \begin{align}
    P = \frac{\rho_0 \rmv_w}{k_\perp\omega} \left[\omega^2 f_{sc}^2 + \left(\omega^2-f^2\right)^2\frac{k_z^2}{k_\perp^2}\right]^{\frac{1}{2}}.
  \end{align}
\end{center}

\noindent Note however, the pressure matching condition fails for these modes at the pole for sub-inertial waves
($\omega\rightarrow f$). The reason is that the propagation domain of sub-inertial GIWs excludes the pole and it becomes
increasingly concentrated toward the equator for faster rotation rates \citep[e.g.,][]{dintrans00,prat16}. Using the
dispersion relationship, and equating the two pressures, yields the following equation for the vertical wave velocity

\vspace{-0.25truein}
\begin{center}
  \begin{align}
    \rmv_w = \omega k_\perp \rmv^2 \left[\omega^2 f_s^2 + \left(N_R^2-\omega^2\right)\left(\omega^2-f^2\right)\right]^{-\frac{1}{2}},
  \end{align}
\end{center}

\noindent Therefore, the wave energy flux density becomes

\vspace{-0.25truein}
\begin{center}
  \begin{align}
    F_z=\frac{1}{2}\frac{\rho_0 \rmv^4 k_\perp \omega\left[\omega^2 f_{ss}^2 + \left(N_R^2-\omega^2\right)\left(\omega^2-f^2\right)\right]^{\frac{1}{2}}}{\omega^2 f_s^2 + \left(N_R^2-\omega^2\right)\left(\omega^2-f^2\right)}.
  \end{align}
\end{center}

Flows in a gravitationally stratified convective medium tend to have an extent in the direction of gravity that is much
larger than their extent in the transverse directions.  Therefore, the horizontal wavenumber of the convective flows is
much greater than the vertical wavenumber. This implies that $k_{\perp,c} \approx \omega_c/\rmv$. For efficient wave
excitation, the frequency of the wave needs to be close to the source frequency \citep{press81,lecoanet13}, which means
that the horizontal scale of the waves will be similar to that of the convection. More generally, there will be a
distribution of excitation efficiency as a function of the wave frequency $\omega$, which may be peaked near the
convective overturning frequency $\omega_c$.  However, since this distribution is unknown, the full frequency dependence
is retained.  This assumption simplifies the wave energy flux to

\vspace{-0.25truein}
\begin{center}
  \begin{align}
    F_z\approx\frac{1}{2}\frac{\rho_0 \rmv^3 \omega^2\left[\omega^2 f_{ss}^2 + \left(N_R^2-\omega^2\right)\left(\omega^2-f^2\right)\right]^{\frac{1}{2}}}{\omega^2 f_s^2 + \left(N_R^2-\omega^2\right)\left(\omega^2-f^2\right)}.
  \end{align}
\end{center}

\noindent In this case, the nonrotating wave energy flux estimate found in \citetalias{press81} can be recovered when
letting $\Omega_0\rightarrow0$ as

\vspace{-0.25truein}
\begin{center}
  \begin{align}
    F_0\approx\frac{\rho_0 \rmv_0^3 \omega}{2\left(N_R^2-\omega^2\right)^{\frac{1}{2}}}.
  \end{align}
\end{center}

Finally, taking the ratio of the two energy fluxes to better isolate the changes induced by rotation, assuming that the
Brunt-V\"{a}is\"{a}l\"{a} frequency is not directly impacted by rotation, and at a fixed wave frequency $\omega$, one
has that

\vspace{-0.25truein}
\begin{center}
  \begin{align}
    \frac{F_z}{F_0} \approx \left(\frac{\rmv}{\rmv_0}\right)^{\!\!3} \!\frac{\omega\!\left\{\!\left(N_R^2\!-\!\omega^2\right)\!\left[\omega^2
    f_{ss}^2 \!+\! \left(N_R^2\!-\!\omega^2\right)\left(\omega^2\!-\! f^2\right)\!\right]\!\right\}^{\!\frac{1}{2}}}{\omega^2
    f_s^2 \!+\! \left(N_R^2\!-\! \omega^2\right)\left(\omega^2\!-\! f^2\right)}.
  \end{align}
\end{center}

\begin{figure}[t!]
  \begin{center}
    \vspace*{1cm}\includegraphics[width=0.45\textwidth]{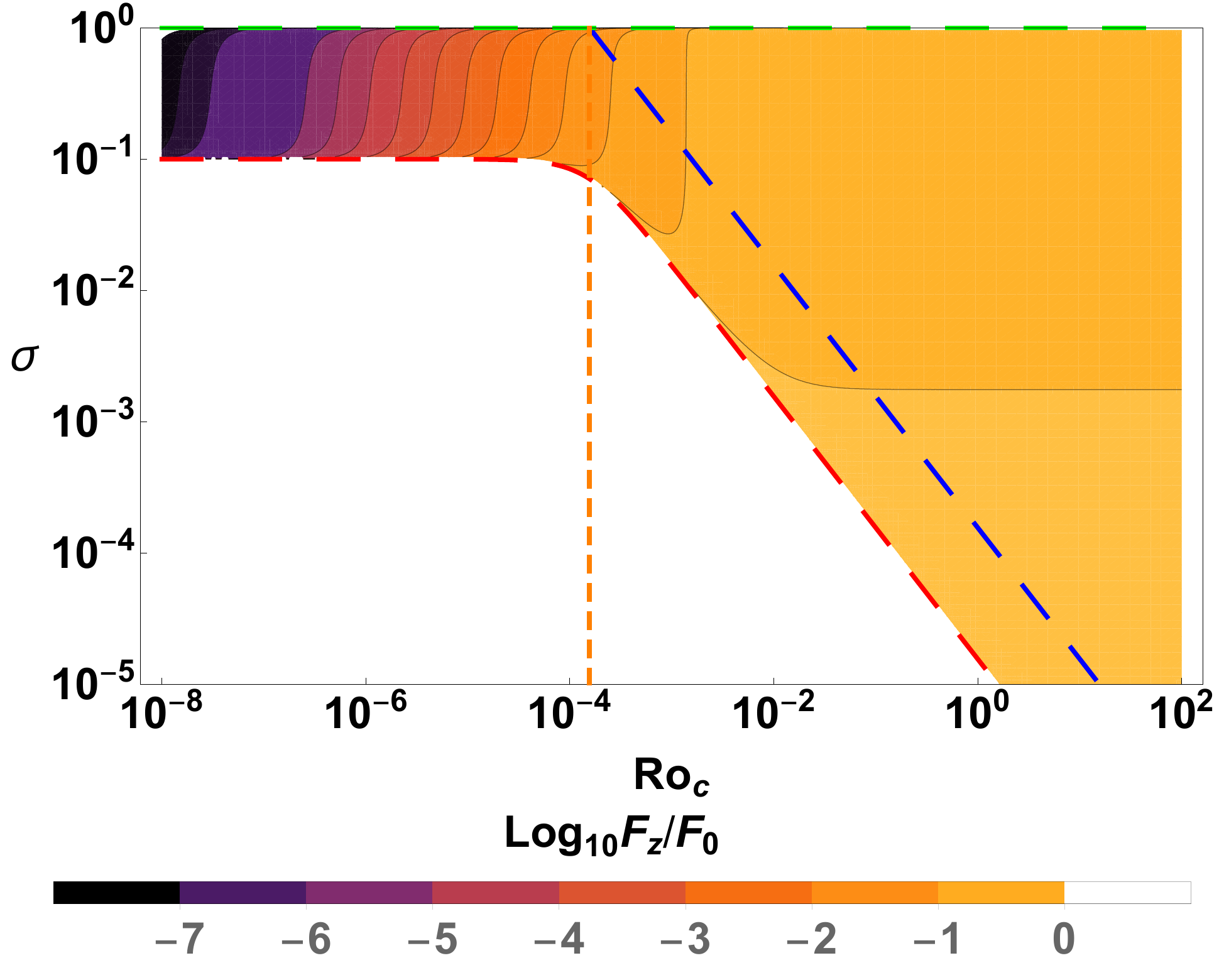} 
    \caption{Convective Rossby number dependence of the ratio of the interfacial gravito-inertial wave energy flux
      excited by rotating convection relative to the nonrotating case $F_z/F_0$ for the nondiffusive convection model
      near the equator, with an interface stiffness of $S = 10^{3}$ and a horizontal direction of
      $\psi=\pi/2$. The red dashed line indicates the lower frequency cutoff $\sigma_{-}$, the green dashed line
      indicates the upper cutoff frequency of $\sigma=1$ since the wave energy flux is being compared to a non-rotating
      case, whereas the blue dashed line indicates a wave Rossby number of $\Rowt=1$.  The vertical dashed orange line
      indicates the critical convective Rossby number. }\label{fig:wave_scaling}
  \end{center}
\end{figure}

\begin{figure}[t!]
  \begin{center}
    \vspace*{1cm}\includegraphics[width=0.45\textwidth]{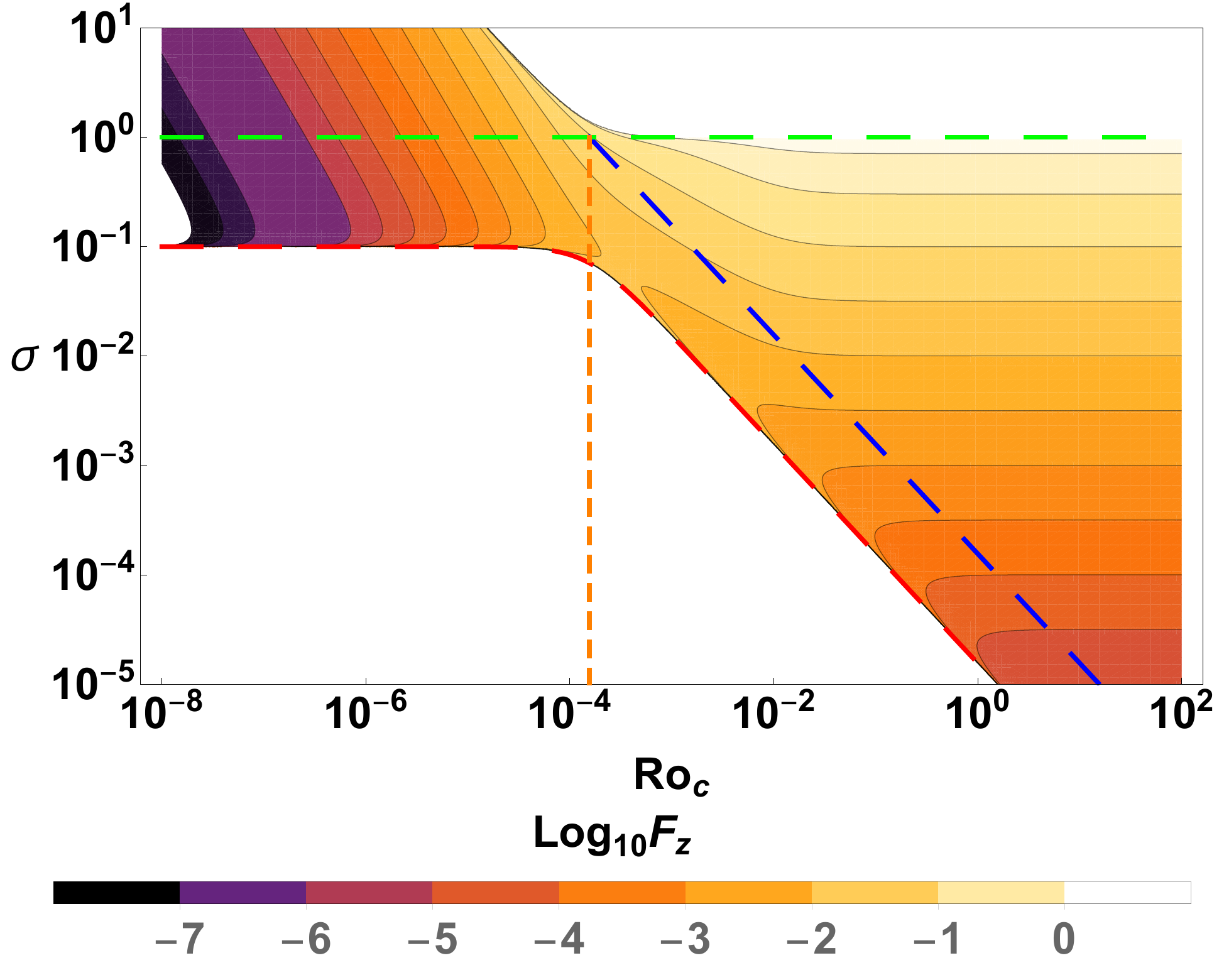} 
    \caption{Convective Rossby number dependence of the interfacial gravito-inertial wave energy flux normalized by the
      non-rotating convective flux excited by rotating convection $F_z$ for the nondiffusive convection model, with parameters as in
      Figure \ref{fig:wave_scaling}, showing the scaling of the flux for $\sigma>1$.}\label{fig:wave_scalingb}
  \end{center}
\end{figure}

\noindent To make this a bit more parametrically tractable, one can normalize the wave frequency as $\sigma=\omega / N_R$, 
and cast the rotational terms into a product of the stiffness of the transition $S=N_R/N_0$, with the convective Rossby 
number of the convection zone as defined above in \S \ref{sec:genframe} with Equations \ref{eqn:qdef} and \ref{eqn:rossby}. 
Doing so yields

\vspace{-0.25truein}
\begin{center}
  \begin{align}
    &\frac{F_z}{F_0}\! \approx\! \left(\frac{\rmv}{\rmv_0}\right)^{\!\!3} \!\left[\Ro_{\mathrm{w}}^{-2}\sin^2{\!\theta}\!+\! \left(\sigma^{-2}\!-\!1\right)\!\!\left(\!1\!-\!\Ro_{\mathrm{w}}^{-2}\cos^2{\!\theta}\!\right)\!\right]^{\!-1}\nonumber\\
    &\left[\!\left(\sigma^{-2}\!-\!1\right)\!\Ro_{\mathrm{w}}^{-2}\sin^2{\!\theta}\sin^2{\!\psi}\!+\!\left(\sigma^{-2}\!-\!1\right)^2\!\left(1\!-\!\Ro_{\mathrm{w}}^{-2}\cos^2{\!\theta}\right)\!\right]^{\!\frac{1}{2}}\!\!\!,
  \end{align}
\end{center}

\noindent where $\Row=\omega/2\Omega_0=5\pi\sigma \Roc S/\sqrt{6}$ and the wave Rossby number is
$\Rowt=\Row/\sqrt{\sin^2{\theta}\sin^2{\psi}+\cos^2{\theta}}$.

\begin{figure}[t!]
  \begin{center}
    \includegraphics[width=0.5\textwidth]{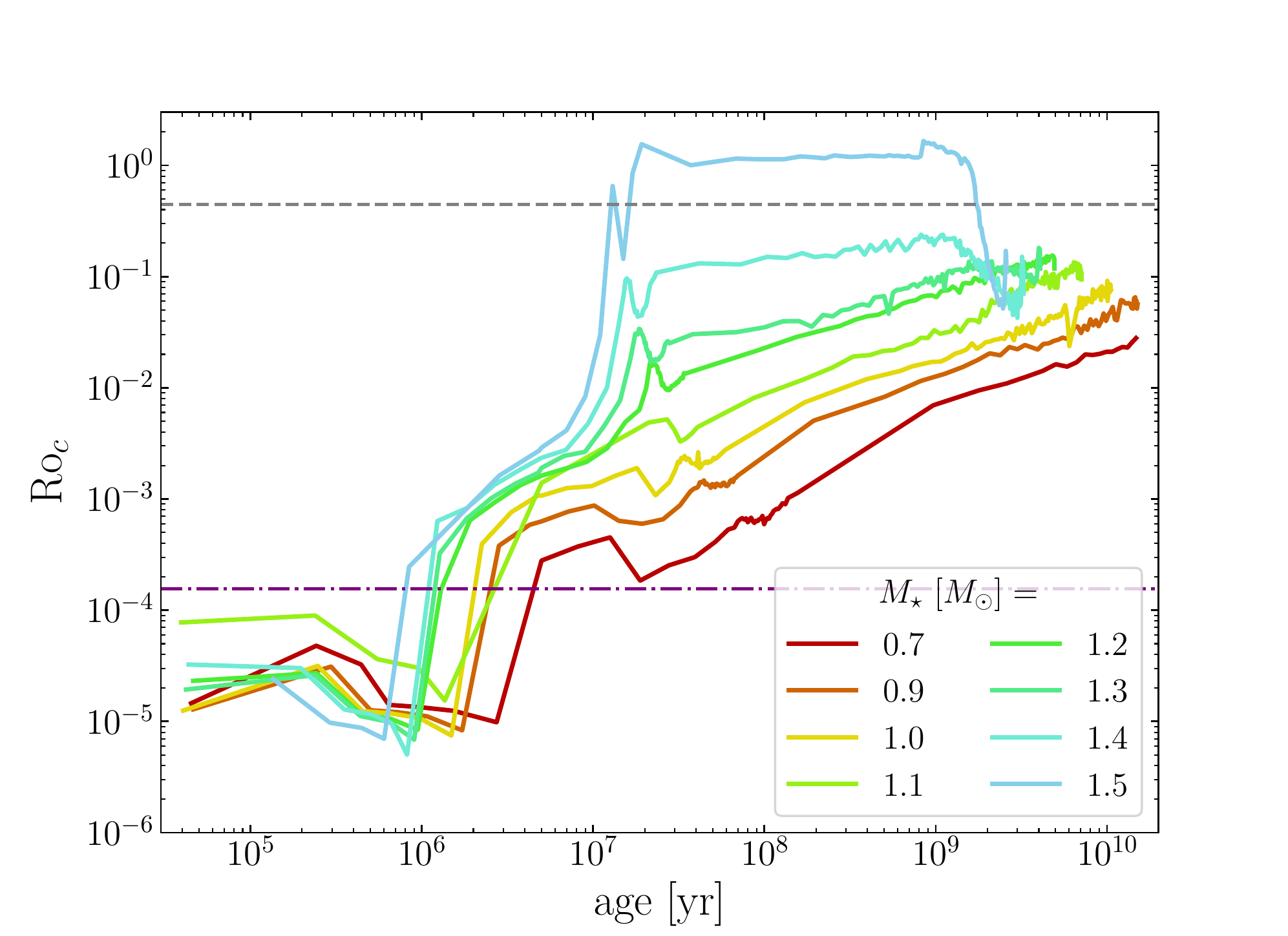} 
    \caption{Variation of the convective Rossby number at the base of the convective envelope of low-mass stars (from
      0.7 to 1.5 $M_{\odot}$) along their evolution. The critical convective Rossby numbers are shown for which a
      potential increase of the excitation rate of GIWs (when compared to the one of pure IGWs) is expected. The purple
      dashed-dot line corresponds to the case of their interfacial excitation and the dashed gray line to the case of
      the excitation triggered by Reynolds stresses.}\label{fig:evolution}
  \end{center}
\end{figure}

This is depicted in Figures \ref{fig:wave_scaling} and \ref{fig:wave_scalingb}, where the colored region exhibits the
magnitude of the logarithm of the energy flux ratio and the energy flux itself. An interfacial stiffness of $S=10^3$ is
chosen as it is a rough estimate of the potential stiffness in most stars, being the ratio of the buoyancy time-scale in
the stable region to the convective overturning time. The choice of latitude determines the width of the frequency band
of sub-inertial waves, where it is a minimum at the pole and maximum near the equator.  This is due to the presence of a critical latitude of the gravito-inertial waves, where sub-inertial waves become evanescent ($\cos^2{\theta_c} = \Row^2$). The direction of $\psi=\pm\pi/2$
is chosen as it represents the maximum value of the energy flux ratio for the choice of other parameters and represents
the waves traveling toward either of the poles as the energy flux ratio is an even parity function of the horizontal
direction. Specifically, the poleward wave energy flux ratio is greater than the other extremal choice of the prograde
or retrograde wave energy flux ratios. In particular, given the range of $\omega_{\pm}$, there are no sub-inertial waves
in the prograde or retrograde propagation case ($\psi=\{0,\pi\}$, respectively), whereas the super-inertial waves may
still propagate with roughly the same frequency range.  The white region corresponds to the domain of evanescent waves
for a given convective Rossby number with frequencies below the lower cut-off frequency ($\sigma_{-}$, dashed red line)
for propagating GIWs. At frequencies above this threshold there is a frequency dependence of the energy flux ratio until
reaching the upper cut-off where $\sigma=\omega/N_R = 1$, which arises due to the domain of validity when comparing GIW
to gravity wave energy fluxes. Indeed, gravity waves may propagate if $\omega<N_R$, whereas super-inertial GIWs may
propagate even when $N_R<\omega<\omega_{+}$. The transition between super-inertial and sub-inertial waves is demarked
with the dashed blue line, with super-inertial waves for $\Rowt>1$ and sub-inertial waves for $\Rowt<1$. Here,
interfacially-excited super-inertial waves exhibit both a frequency and convective Rossby number
dependence. Specifically, the wave energy flux decreases algebraically with frequency at a fixed convective Rossby
number and have a reduced energy flux for convective Rossby numbers below unity.  The interfacially-excited sub-inertial
waves possess a small frequency domain at a fixed convective Rossby number over which they are propagative.  The
sub-inertial wave energy flux increases with decreasing convective Rossby number until a critical convective Rossby
number $\Ro_{c,\mathrm{crit}} = \sqrt{6}/\left(5\pi S\right)$ as depicted by the vertical dashed orange line in Figure
\ref{fig:wave_scaling}. Below this critical convective Rossby number, the sub-inertial wave energy flux decreases and
their frequency domain is further restricted until it vanishes entirely and there are no propagative super-inertial
waves .  The effect of the stiffness is to lower (raise) the value of the critical convective Rossby number for larger
(smaller) values of $S$, which corresponds to the ratio of the buoyancy time-scale in the radiative zone to the
convective overturning time. This may have important consequences for the wave-induced transport of angular momentum
during the evolution of rotating stars.  In particular, the convective Rossby number can vary by several orders of
magnitude over a star's evolution from the PMS to its ultimate demise
\citep[e.g.][]{landin10,mathis16,charbonnel17}. Moreover, it can vary internally as a function of radius due to the
local amplitude of the convective velocity and due to transport processes, angular momentum loss through winds, and
structural changes that modify the local rotation rate \citep{mathis16}. Figure \ref{fig:evolution} presents the
variation of the convective Rossby number at the base of the convective envelope of low-mass stars (from 0.7 to 1.5
M$_{\odot}$) throughout their evolution. These convective Rossby numbers have been calculated using grids of stellar
models that take into account rotation computed with the STAREVOL code
\citep{Siessetal2000,Palaciosetal2003,Decressinetal2009,Amardetal2016}. The details of the micro- and macro-physics used
for these grids are described in \citet{Amardetal2019}. The dot-dashed purple line provides the value of the critical
convective Rossby number ($\Ro_{c,\mathrm{crit}} = \sqrt{6}/\left(5\pi S\right)$, shown here for $S=10^3$, for which an
increase of the interfacial excitation of GIWs can be expected. For all the stars considered here, which have a median
initial rotation (i.e. 4.5 days), this should happen during their PMS.

\begin{figure}[t!]
  \begin{center}
    \vspace*{1cm}\includegraphics[width=0.45\textwidth]{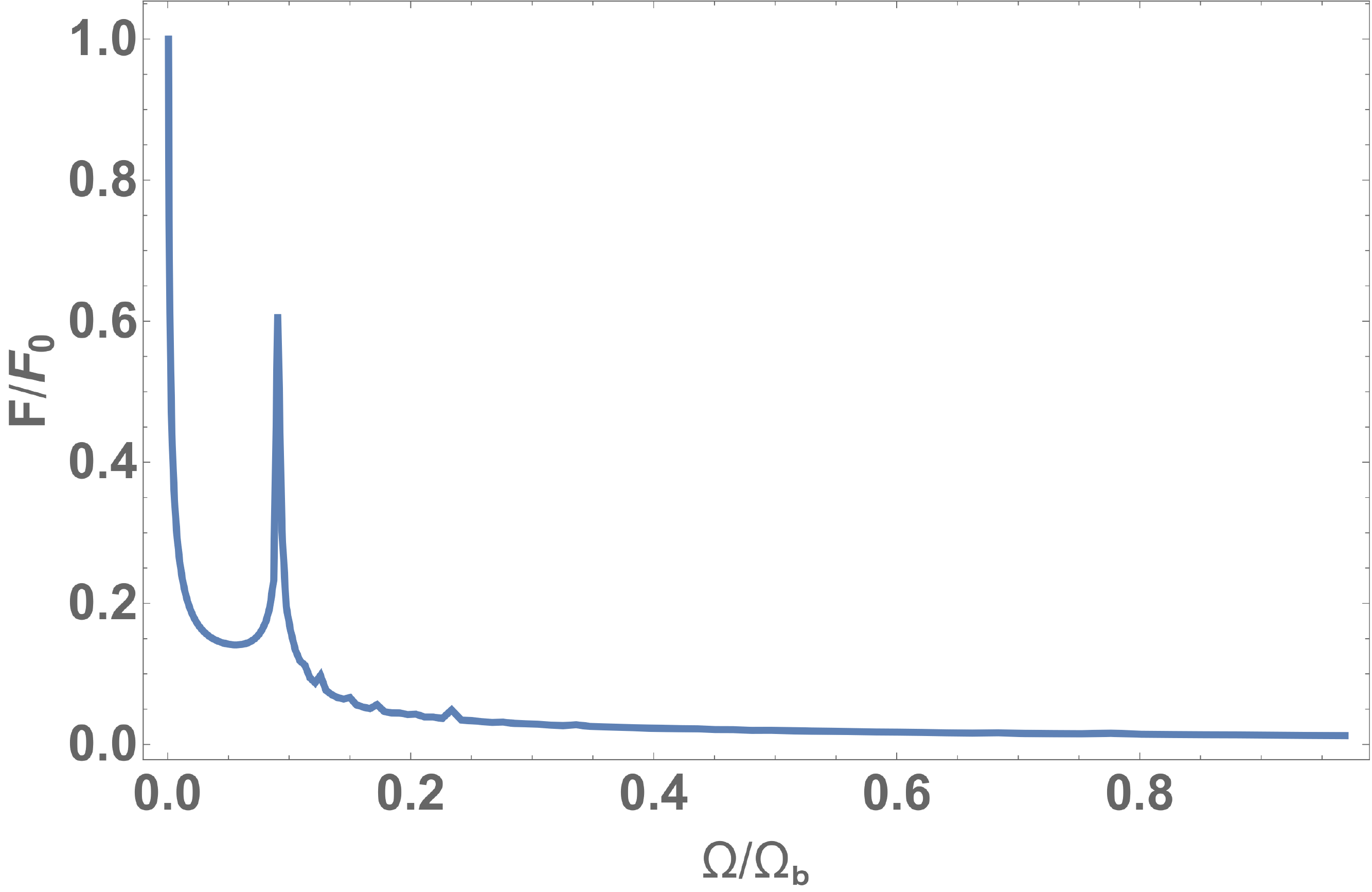} 
    \caption{Scaling of the relative flux integrated over $\theta$, $\psi$, and frequency with respect to the ratio of
      the rotation rate $\Omega$ to the break-up rotation rate $\Omega_{\mathrm{b}}$, with the stiffness $S=10^3$ and a
      minimum Rossby number $\Ro_{\mathrm{b}}=10^{-5}$ at $\Omega_{\mathrm{b}}$, showing the peak at
      $\Omega/\Omega_{\mathrm{b}}\approx 0.091$.}\label{fig:intint_scaling}
  \end{center}
\end{figure}

The flux ratio $F_z/F_0$ integrated over latitude $\theta$, propagation direction $\psi$, and frequency is shown in
Figure \ref{fig:intint_scaling}.  This illustrates the general rotational trend of the interfacial flux, namely that it
decreases with increasing rotation rate. However, there is a peak at a rotational frequency that depends upon the choice
of stiffness $S$ and the Rossby number of the convection at the breakup velocity $\Ro_{\mathrm{b}}$. Thus, for stars
with a modest rotation rate below approximately $0.2 \Omega_b$, the interfacial or pressure-driven GIW wave flux could
play a role in transport processes that is at least as important as the transport by IGWs.  Yet, for more rapidly
rotating stars, this flux becomes fairly negligible due primarily to the reduction in the convective velocity
amplitudes. Nevertheless, given the complex and nonanalytic form of the full integral, the exploration of the parameter
dependence of this peak will be left for future work. As a means of comparison, consider the spherical Couette flow
laboratory experiments of \citet{hoff16}, where it is found that the kinetic energy of the dominant inertial mode
increases with decreasing wave Rossby number. Below a critical wave Rossby number this leads to a wave breaking and an
increase of small-scale structures at a critical Rossby number, which may be similar to the large increase of wave
energy flux for sub-inertial waves below the critical convective Rossby number described above.

\section{Reynolds Stress Contributions to GIW Amplitudes}\label{sec:reynoldswaves}

As a means of comparison, the amplitude and the wave energy flux of the GIWs may be computed exactly when using the
convection model presented earlier, where the impact of rotation on the waves is treated coherently.  In a means similar
to \citet{goldreich90} and \citet{lecoanet13}, although with a greater degree of computational complexity, one may
derive the wave amplitudes for GIWs in a f-plane. As seen in \citet{mathis14}, one must first find solutions to the
homogeneous Poincar\'{e} equation for the GIWs and then use linear combinations of those solutions to construct
solutions to the forced equation in the convection zone. These equations result from writing the linearized equations of
motion in a f-plane as a single equation for the vertical velocity $W$ as

\vspace{-0.25truein}
\begin{center}
  \begin{align}
    \left[\partial_{tt}\nabla^2 + 4\left(\bomega\cnabla\right)^2  + N^2\nabla_\perp^2\right]W = \partial_t\mathbb{S}, \label{eqn:poincare}
  \end{align}
\end{center}

\noindent where $\mathbb{S}$ is the convective source term described in detail below. Note that the thermal sources
derived in \citet{samadi01} have been neglected here as in \citet{mathis14}, for they have been found to be
comparatively small for gravity waves when compared to Reynolds stresses \citep{belkacem09a}. In addition, the damping
mechanisms (i.e. the radiative damping and the damping due to convection-wave interactions) are neglected here. As
pointed out in \citet{samadi15}, the value of the amplitude of stochastically excited waves is proportional to the
ratio of the energy injection rate that measures the efficiency of the couplings of turbulent motions with waves and of
the damping. The focus of this work is on the energy injection rate while getting a coherent treatment of the turbulent
damping of waves in rotating stars will be considered in forthcoming work.

In the stable region, where $\mathbb{S}$ is assumed to vanish, it can be shown that if one follows the methodology of
constructing normal modes as in \citet{gerkema05} then the solutions of the homogeneous Poincar\'{e} equation for GIWs
may be expanded as $w = \mathrm{w}(\chi,z) e^{i\omega t}$. The horizontal coordinate $\chi=\cos{\psi}x+\sin{\psi}y$
corresponds to the distance along the direction of the wave propagation with an angle $\psi$ in the horizontal plane as
seen in \citet{gerkema05} and Figure 1 of \citet{mathis14}.  Therefore, as before, the solution of the forced
Poincar\'{e} equation for GIWs in the convection zone may be expanded as

\vspace{-0.25truein}
\begin{center}
  \begin{align}
    W\left(\chi,z,t\right) &= \sum_nA_n\left(t\right)\mathrm{w}_n\left(\chi,z\right) e^{i\omega t}\\
    &= \sum_{n}A_n\left(t\right) \psi_n\left(z\right) e^{i k_n\left(\chi + \delta z\right) + i\omega t},
  \end{align}
\end{center}

\noindent where $\omega$ is the chosen frequency, with $k_n$ being the sequence of eigenvalues associated with
it, the $\psi_n$ are the eigenmodes of the reduced Poincar\'{e} equation (see Equation \ref{eqn:poinred}), and $A_n$ is
its amplitude. Technically, the full velocity field would be an integral over all frequencies and the sum over modes
associated with each frequency. However, for simplicity, this discussion will at first focus on a single frequency taken
to be within the band of propagative frequencies. Substituting this into Equation \ref{eqn:poincare}, yields an equation
for the mode amplitudes given a source function

\vspace{-0.25truein}
\begin{center}
  \begin{align}
    &\sum_n\left\{\left[\partial_{tt}A_n + 2 i \omega\partial_tA_n\right]\left[\partial_{zz}-k_n^2\right]
    + A_n\left[\left(f^2-\omega^2\right)\partial_{zz} \right.\right.\nonumber\\
    &\left.\left.+ 2 i f f_{ss}k_n\partial_z
      + k_n^2\left(\omega^2-N^2-f_{ss}^2\right)\right]\right\}\mathrm{w}_n e^{i\omega t} = \partial_t\mathbb{S}. \label{eqn:poincareexp}
  \end{align}
\end{center}

\noindent Noting that the second term is simply the homogeneous equation, it vanishes, leaving

\vspace{-0.25truein}
\begin{center}
  \begin{align}
    \sum_n&\left[\partial_{tt}A_n + 2 i \omega\partial_tA_n\right]\nonumber\\
    &\times\left[\partial_{zz}\psi_n+2ik_n\delta\partial_z\psi_n-k_n^2\left(\delta^2+1\right)\psi_n\right]\nonumber\\
    &\times e^{ik_n\left(\chi+\delta z\right)+i\omega t} = \partial_t\mathbb{S}.
  \end{align}
\end{center}

\noindent Utilizing Equation \ref{eqn:poinred}, this becomes

\vspace{-0.25truein}
\begin{center}
  \begin{align}
    &\sum_n\left[\partial_{tt}A_n + 2 i \omega\partial_tA_n\right]\nonumber\\
    &\times\!\!\left[2ik_n\delta\partial_z\psi_n
      -k_n^2\!\!\left(\!\!\frac{f^2-N^2}{f^2-\omega^2}+\frac{\left(\omega^2+f^2\right)f_{ss}^2}{\left(f^2-\omega^2\right)^2}\!\!\right)\!\!\psi_n\right]\nonumber\\
    &\times e^{ik_n\left(\chi+\delta z\right)+i\omega t} = \partial_t\mathbb{S}.
  \end{align}
\end{center}

\noindent Assuming homogeneous Dirichlet boundary conditions on $\psi_n$ and that the change in the amplitudes at
infinity are zero, with an initial condition of being zero, this can be integrated against a single conjugate mode of
index $m$ to see that the constant amplitude is

\vspace{-0.25truein}
\begin{center}
  \begin{align}
    \langle A_n\rangle &=
    \frac{i\int_0^Ldz\int_{-\infty}^{\infty}d\chi\int_{-\infty}^{\infty}dt\partial_t\mathbb{S}\psi_n^* e^{-ik_n\left(\chi+\delta z\right)-i\omega t}}{
      2\omega k_n^2\int_0^Ldz\left(\frac{f^2-N^2}{f^2-\omega^2}+\frac{\left(\omega^2+f^2\right)f_{ss}^2}{\left(f^2-\omega^2\right)^2}\right)\left|\psi_n\right|^2}, \label{eqn:amps}
  \end{align}
\end{center}

\noindent where the normalization $c_n$ follows from the orthogonality condition on the $\psi_n$,

\vspace{-0.25truein}
\begin{center}
  \begin{align}
    c_n^2 \delta_{nm}= \frac{1}{L}\int_0^Ldz\left(\frac{\omega^2-N^2}{f^2-\omega^2}+\frac{\omega^2f_{ss}^2}{\left(f^2-\omega^2\right)^2}\right)\psi_m^*\psi_n.
  \end{align}
\end{center}

\noindent where $L = z_2-z_1$ is the depth of the domain. The convective source term is

\vspace{-0.25truein}
\begin{center}
  \begin{align}
    \mathbb{S}&=\partial_z\dvg{\mathbf{F}}-\nabla^2F_z = \partial_z\grad_\perp\bcdot \mathbf{F} -\nabla_\perp^2F_z,
  \end{align}
\end{center}

\noindent where $\mathbf{F}=\dvg{\left(\boldsymbol{\rmv}\otimes\boldsymbol{\rmv}\right)}$ are the Reynolds stresses
due to the convective velocities $\boldsymbol{\rmv}$.  This can be further simplified noting the definition of the perpendicular
direction, yielding

\vspace{-0.25truein}
\begin{center}
  \begin{align}
    \mathbb{S}&= \partial_{\chi z}F_\chi - \partial_{\chi}^2F_z, \\
    &=\partial_{\chi\chi z}\left(\rmv_\chi^2-\rmv_z^2\right) + \left(\partial_{\chi zz}-\partial_{\chi\chi\chi}\right)\rmv_\chi\rmv_z.
  \end{align}
\end{center}

The integral in the numerator of Equation \ref{eqn:amps} can be identified as a Fourier transform of the source in time and
space. Treating it as such, it becomes

\vspace{-0.25truein}
\begin{center}
  \begin{align}
    &\int_0^Ldz\int_{-\infty}^{\infty}d\chi\int_{-\infty}^{\infty}dt\partial_t\mathbb{S}\psi_n^* e^{-ik_n\left(\chi+\delta z\right)-i\omega t} \nonumber\\
    &=\!\omega\!\!\int_0^L\!\!dz\!\left[i k_n^2\partial_z\!\left(\widetilde{\rmv_z^2\,}\!-\!\widetilde{\rmv_\chi^2}\right) \!-\! k_n \!\left(\partial_{zz}\!+\!k_n^2\right)\!\widetilde{\rmv_\chi\rmv_z}\right]\!\psi_n^* e^{-ik_n\delta z}.
  \end{align}
\end{center}

\noindent In turn this is a Fourier transform of a product, or a convolution in spectral space of the Reynolds stress
with a Heaviside function $H$ that confines the convection to a the convective region and the reduced eigenmodes.  Under
this approach, the previous equation yields

\vspace{-0.25truein}
\begin{center}
  \begin{align}
    &\omega \int_{-\infty}^{\infty}dk'\left[k'k_n^2\left(\widehat{\rmv_z^2\,}\left(k'\right)-\widehat{\rmv_\chi^2}\left(k'\right)\right)\right.\nonumber\\
      &\left.+ k_n\left(k'^2-k_n^2\right)\widehat{\rmv_\chi\rmv_z}\left(k'\right)\right]\widehat{H \psi_n^*}\left(k_n\delta-k'\right).
  \end{align}
\end{center}

Assuming henceforth that the Brunt-V\"{a}is\"{a}l\"{a} frequency is a discontinuous jump of an amplitude $N = S
\omega_c$, there is an exact solution for all three wave classes, sub-inertial, inertial, and super-inertial. This
assumption provides an approximation of the stratification in a star, but captures its order of magnitude effects. This
means that all integrals except the one of the Reynolds stresses can be evaluated. The latter depends upon the
turbulence model that is chosen. The one introduced at the beginning of this paper will be examined here. Specifically,
with this choice of $N$, the reduced Poincar\'{e} equation becomes

\vspace{-0.25truein}
\begin{center}
  \begin{align}
    \left\{\begin{array}{lr}
    \partial_{zz}\psi_n + k_n^2\alpha^2\psi_n = 0 & 0\le z < \ell_s\\
    \partial_{zz}\psi_n + k_n^2\beta^2\psi_n = 0 & \ell_s\le z\le L
    \end{array}\right.,
  \end{align}
\end{center}

\noindent where

\vspace{-0.25truein}
\begin{center}
  \begin{align}
    \alpha^2 &= \frac{N^2-\omega^2}{\omega^2-f^2}+\frac{\omega^2f_{ss}^2}{\left(\omega^2-f^2\right)^2},\\
    \beta^2 &= \frac{\omega^2f_{ss}^2}{\left(\omega^2-f^2\right)^2}-\frac{\omega_c^2+\omega^2}{\omega^2-f^2},
  \end{align}
\end{center}

\noindent and where $\omega_c$ is the convective overturning time and $\ell_s = z_c-z_1$ is the depth of the radiative-convective
interface. The boundary conditions are that $\psi_n(0) = \psi_n(L) = 0$, and with matching conditions and momentum
continuity at the interface leading to the dispersion relationship. With these choices, above equations admit the
following solutions for the sub-inertial waves

\vspace{-0.25truein}
\begin{center}
  \begin{align}
    \psi_n = \left\{\begin{array}{lr}
    -\displaystyle{\frac{\sin\left(k_n\beta \left(L-\ell_s\right)\right)}{\cos\left(k_n\beta L\right) \sin\left(k_n\alpha \ell_s\right)}\sin\left(k_n\alpha z\right)} & 0\le z<\ell_s\\
    \sin\left(k_n\beta z\right) - \tan\left(k_n\beta L\right)\cos\left(k_n\beta z\right) & \ell_s\le z\le L
    \end{array}\right.,
  \end{align}
\end{center}

\noindent with a dispersion relationship

\vspace{-0.25truein}
\begin{center}
  \begin{align}
    \alpha\tan\left[k_n\beta\left(L-\ell_s\right)\right] + \beta\tan\left[k_n\alpha\ell_s\right] = 0.
  \end{align}
\end{center}

Similarly, the super-inertial waves are

\vspace{-0.25truein}
\begin{center}
  \begin{align}
    \psi_n = \left\{\begin{array}{lr}
    -\displaystyle{\frac{\sinh\left(k_n\beta \left(L-\ell_s\right)\right)}{\cosh\left(k_n\beta L\right) \sin\left(k_n\alpha \ell_s\right)}\sin\left(k_n\alpha z\right)} & 0\le z<\ell_s\\
    \sinh\left(k_n\beta z\right) - \tanh\left(k_n\beta L\right)\cosh\left(k_n\beta z\right) & \ell_s\le z\le L
    \end{array}\right.,
  \end{align}
\end{center}

\noindent with a dispersion relationship

\vspace{-0.25truein}
\begin{center}
  \begin{align}
    \alpha\tanh\left[k_n\beta\left(L-\ell_s\right)\right] + \beta\tan\left[k_n\alpha\ell_s\right] = 0.
  \end{align}
\end{center}

\noindent Note that with $\sin\left(k_n\alpha \ell_s\right)$ in the numerator, there are certain values of $k_n \alpha
\ell_s = m \pi$ with $m$ some integer where this solution is invalid.  This provides an additional selection criterion
on the values of S that have solutions.   Note that the inertial waves already are normalized with $c_n=1$. The integrals for
the denominator in Equation \ref{eqn:amps} are very similar.

Finally, from Equation 62 in \citet{andre17}, the vertical wave flux
for a single mode is given as

\vspace{-0.25truein}
\begin{center}
  \begin{align}
    F_z = \frac{\rho_0}{2}\left(\frac{f^2-\omega^2}{\omega k_n^2}\right)\left|A_n^2\psi_n\partial_z\psi_n\right|.\label{eqn:flux_def}
  \end{align}
\end{center}

\noindent Note that this definition of the flux is slightly different from the interfacial flux, which used an
approximation of the pressure. Moreover, that interfacial flux is a local model with driving taking place only at the
interface, whereas the current model assesses wave driving throughout the convective zone. The definition of the flux
given in Equation \ref{eqn:flux_def} is consistent with previous studies of gravity wave driving in the bulk of
convective regions \citep[e.g.,][]{press81,goldreich90,lecoanet13}, where it is seen that the flux for a discontinuous
Brunt-V\"{a}is\"{a}l\"{a} frequency is $F_z \approx F_c S^{-1}$, where $F_c$ is the convective flux. This scaling can be
obtained using Equation \ref{eqn:flux_def} if one considers the regime of low-frequency gravity waves in the
non-rotating case (where $f=0$), which are described within the JWKB approximation, assuming that their horizontal
velocity is $v_h\approx v_c=\omega_c/k_c$, where $v_c$, $\omega_c$, and $k_c$ are the convective velocity, frequency,
and wavevector, respectively, while one also sets $\omega\approx\omega_c$ and $k\approx k_c$. Note that within these
assumptions the influence of the spatial behaviour of the eigenmodes is not taken into account.

Thus, with the definition of the amplitude, the flux is

\vspace{-0.25truein}
\begin{center}
  \begin{align}
    &F_z =\frac{\rho_0\left(f^2-\omega^2\right)\left|\psi_n\partial_z\psi_n\right|}{8\omega^3 k_n^6}\nonumber\\
    &\times\frac{\left|\int_0^Ldz\int_{-\infty}^{\infty}d\chi\int_{-\infty}^{\infty}dt\partial_t\mathbb{S}\psi_n^* e^{-ik_n\left(\chi+\delta z\right)-i\omega t}\right|^2}{
      \left(\int_0^Ldz\displaystyle{\left(\frac{f^2-N^2}{f^2-\omega^2}+\frac{\left(\omega^2+f^2\right)f_{ss}^2}{\left(f^2-\omega^2\right)^2}\right)}\left|\psi_n\right|^2\right)^2}. \label{eqn:flux_expand}
  \end{align}
\end{center}

\begin{figure*}[t!]
  \begin{center}
    \includegraphics[width=\textwidth]{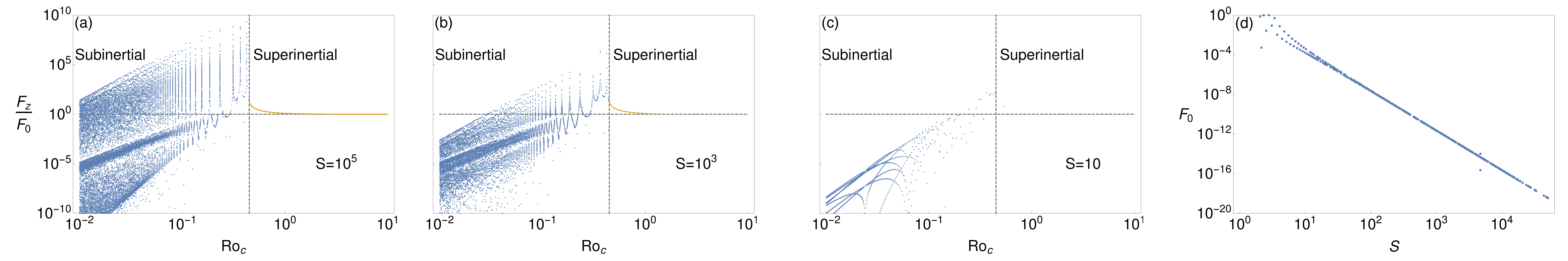} 
    \caption{Scaling of the gravito-inertial wave flux $F_z$, normalized by the gravity wave flux for the non-rotating
      case $F_0$, when excited by columnar convection at the equator for waves, where the stiffness is the ratio
      Brunt-V\"{a}is\"{a}l\"{a} frequency to the rotation frequency $S = N_R/N_0$ is taken to be (a) $10^5$, (b) $10^3$,
      and (c) $10$ and where $\ell_0=\ell_s=L/2$, and $\psi=\pi/2$. The vertical dashed line denotes the transition
      between sub-inertial and super-inertial waves and the horizontal line denotes unity. (d) illustrates the scaling
      of the pure gravity wave flux ($F_0$) normalized by the total convective flux with the stiffness parameter,
      showing that the wave flux is always below the convective flux, but that the gravito-inertial wave flux can be
      greatly amplified in comparison. Note that such mode amplification of GIWs has also been seen in a global model
      \citep[see Figure 7][]{neiner20}}\label{fig:conv_waves}
  \end{center}
\end{figure*}

\noindent Averaging over the stable region, this becomes

\vspace{-0.25truein}
\begin{center}
  \begin{align}
    &F_z =\frac{\rho_0\left(f^2-\omega^2\right)\psi_n^2(\ell_s)}{8\omega^3 k_n^6 \ell_s}\nonumber\\
    &\times\frac{\left|\int_0^Ldz\int_{-\infty}^{\infty}d\chi\int_{-\infty}^{\infty}dt\partial_t\mathbb{S}\psi_n^* e^{-ik_n\left(\chi+\delta z\right)-i\omega t}\right|^2}{
      \left(\int_0^Ldz\displaystyle{\left(\frac{f^2-N^2}{f^2-\omega^2}+\frac{\left(\omega^2+f^2\right)f_{ss}^2}{\left(f^2-\omega^2\right)^2}\right)}
      \left|\psi_n\right|^2\right)^2}. \label{eqn:flux}
  \end{align}
\end{center}

The integral in the denominator of the wave flux for the sub-inertial and super-inertial waves are thus

\vspace{-0.25truein}
\begin{center}
  \begin{align}
    &D_n = \sec^4\left(k_n L\beta\right)\nonumber\\
    &\times\!\!\left[\frac{\sin^2\!\left(k_n\beta\ell_0\right)}{\sin^2\left(k_n\alpha\ell_s\right)}\!
      \left(\!\frac{f^2-N^2}{f^2-\omega^2}+\frac{\left(\omega^2+f^2\right)f_{ss}^2}{\left(f^2-\omega^2\right)^2}\!\right)\right.\nonumber\\
      &\qquad\times\left(\frac{\ell_s}{2}\!-\!\frac{\sin\left(2k_n\alpha\ell_s\right)}{4k_n\alpha}\right)\nonumber\\
      &\left.+\left(\!\frac{f^2+\omega_c^2}{f^2-\omega^2}+\frac{\left(\omega^2+f^2\right)f_{ss}^2}{\left(f^2-\omega^2\right)^2}\!\right)
      \left(\frac{\ell_0}{2}-\frac{\sin\left(2k_n\beta\ell_0\right)}{4k_n\beta}\right)\right]^2\!\!\!\!,
  \end{align}
\end{center}

\noindent for the sub-inertial waves and

\vspace{-0.25truein}
\begin{center}
  \begin{align}
    &D_n = \mathrm{sech}^4\left(k_n L\beta\right)\nonumber\\
    &\times\!\!\left[\frac{\sinh^2\!\left(k_n\beta\ell_0\right)}{\sin^2\left(k_n\alpha\ell_s\right)}\!
      \left(\!\frac{f^2-N^2}{f^2-\omega^2}+\frac{\left(\omega^2+f^2\right)f_{ss}^2}{\left(f^2-\omega^2\right)^2}\!\right)\right.\nonumber\\
      &\qquad\times\left(\frac{\ell_s}{2}\!-\!\frac{\sin\left(2k_n\alpha\ell_s\right)}{4k_n\alpha}\right)\nonumber\\
      &\left.+\left(\!\frac{f^2+\omega_c^2}{f^2-\omega^2}+\frac{\left(\omega^2+f^2\right)f_{ss}^2}{\left(f^2-\omega^2\right)^2}\!\right)
      \left(\frac{\ell_0}{2}+\frac{\sinh\left(2k_n\beta\ell_0\right)}{4k_n\beta}\right)\right]^2\!\!\!\!,
  \end{align}
\end{center}

\noindent for the super-inertial waves.

The integral in the numerator can be computed exactly for the convection model discussed in Section \ref{sec:genframe}.  Specifically, using the definition of the velocities there, e.g. 

\vspace{-0.25truein}
\begin{center}
    \begin{align}
       \mathrm{v}_z &= \mathrm{v}\left(\Roc\right) \sin{\left(\frac{\pi z}{\ell_0}\right)} e^{\left(ik_\perp \chi+i\omega_c t\right)},\\ 
       \mathrm{v}_\chi &= \frac{i\pi}{\ell_0 k_\perp}\mathrm{v}\left(\Roc\right)\cos{\left(\frac{\pi z}{\ell_0}\right)} e^{\left(ik_\perp \chi+i\omega_c t\right)}, 
    \end{align}
\end{center}

\noindent as follows from the continuity equation for the convection model.  These integrals are

\vspace{-0.25truein}
\begin{center}
  \begin{align}
    &\frac{\pi^4 v^4\sec^2\left(2k_{\perp}L\beta\right)\left(2k_{\perp}^2\ell_0^2-3\pi^2\right)^2}
         {2\ell_0^4\left[\pi^4-2\pi^2k_{\perp}^2\ell_0^2\left(\beta^2+\delta^2\right)+k_{\perp}^4\ell_0^4\left(\beta^2-\delta^2\right)^2\right]^2}\nonumber\\
         &\times\left[\pi^4-2\pi^2k_{\perp}^2\ell_0^2\left(\beta^2+\delta^2\right)+k_{\perp}^4\ell_0^4(\beta^4+14\beta^2\delta^2+\delta^4)\right.\nonumber\\
           &+4\beta\delta k_{\perp}^2\ell_0^2\left(k_{\perp}^2\ell_0^2\left(\beta-\delta\right)^2-\pi^2\right)\nonumber\\
           &\qquad\times\left(\cos\left(2k_{\perp}\ell_0\left(\beta+\delta\right)\right)+\cos\left(2k_{\perp}\ell_0\left(\beta-\delta\right)\right)\right)\nonumber\\
         &\left.-\!\left(\pi^2\!-\!k_{\perp}^2\ell_0^2\left(\beta+\delta\right)^2\right)\!\!
           \left(\pi^2\!-\!k_{\perp}^2\ell_0^2\left(\beta\!-\!\delta\right)^2\right)\!\cos\left(4k_{\perp}\ell_0\beta\right)\right],
  \end{align}
\end{center}

\noindent for the sub-inertial waves where the horizontal and time integrals impose $\omega=2\omega_c$ and
$k_\perp=k_n/2$. For the super-inertial waves, this is

\vspace{-0.25truein}
\begin{center}
  \begin{align}
    &-\frac{\pi^4 v^4\mathrm{sech}^2\left(k_{\perp}L\beta\right)\left(2k_{\perp}^2\ell_0^2-3\pi^2\right)^2}
         {2\ell_0^4\left[\pi^4+2\pi^2k_{\perp}^2\ell_0^2\left(\beta^2-\delta^2\right)+k_{\perp}^4\ell_0^4\left(\beta^2+\delta^2\right)^2\right]^2}\nonumber\\
         &\times\left[\pi^4+2\pi^2k_{\perp}^2\ell_0^2\left(\beta^2-\delta^2\right)+k_{\perp}^4\ell_0^4(\beta^4-14\beta^2\delta^2+\delta^4)\right.\nonumber\\
           &+16\beta^2\delta^2 k_{\perp}^4\ell_0^4\cos\left(2k_{\perp}\ell_0\delta\right)\cosh\left(2k_{\perp}\ell_0\beta\right)\nonumber\\
           &-8\beta\delta k_{\perp}^2\ell_0^2\left(\pi^2+k_{\perp}^2\ell_0^2\left(\beta^2-\delta^2\right)\right)\nonumber\\
           &\qquad\times\sin\left(2k_{\perp}\ell_0\delta\right)\sinh\left(2k_{\perp}\ell_0\beta\right)\nonumber\\
           &-\!\left(k_{\perp}^2\ell_0^2\beta^2+\left(\pi+k_{\perp}\ell_0\delta\right)^2\right)\nonumber\\
           &\left.\qquad\times\left(k_{\perp}^2\ell_0^2\beta^2+\left(\pi-k_{\perp}\ell_0\delta\right)^2\right)
           \!\cosh\left(4k_{\perp}\ell_0\beta\right)\right].
  \end{align}
\end{center}

These waves will attain a maximum flux near the equator, especially for low convective Rossby number where the waves
become increasingly equatorially focused. Thus, evaluating these expressions at the equator, one has a wave flux
analogous to the section on interfacial waves, but excited by the Reynolds stresses in the bulk of the convection
zone. Figure \ref{fig:conv_waves} illustrates this flux for several values of the stiffness, where each value of the
convective Rossby number is computed such that the dispersion relationships are obeyed, leading to its discrete
nature. The sub-inertial waves have an oscillatory character, where some waves achieve a resonance and have a peak in
flux. The peak flux arises at moderate convective Rossby numbers below $1/\sqrt{5}$, due to $\alpha$ being small and
transitioning from super-inertial to sub-inertial waves. The decay of the flux at lower convective Rossby numbers
results from the weakening convective velocities and the increasing horizontal wavenumber of the convection.  The peak
in the super-inertial waves also occurs near $\Ro_c = 1/\sqrt{5}$, above which it decays primarily due to the scaling of
the denominator of the flux, which arises from the hyperbolic trigonometric functions in the structure of the eigenmodes
and it asymptotes to the flux of pure gravity waves driven by nonrotating convection. When considering Figure
\ref{fig:evolution} and also Figure 4 in \citet{mathis16}, where $\Roc =1/\sqrt{5}$ is distinguished by the dashed gray line.
One can see that these phenomena may occur for a majority of low-mass stars along their evolution, in particular during the
PMS (or close to the base of their convective envelope) because of the low values of $\Ro_c$ during these evolutionary
phases (and in these regions).

The actual value of the both the nonrotating and rotating fluxes are very dependent upon the value of the stiffness
chosen due to the dependence on the average of the eigenfunction in the stable region in the numerator and the
normalization in the denominator of the flux derived above.  The non-rotating wave flux normalized by the total
convective flux is shown in Figure \ref{fig:conv_waves}(d). Note that this flux $F_0$ differs from that of Section
\ref{sec:interfacewaves} because the flux defined in Equation \ref{eqn:flux_def} has a complex spatial dependence.  When
averaged over the stable region, this yields $F_0\propto Q(S) S^{-4}$ where $Q(S)$ is the dependence arising from the
integral of the source term and the average value of $|\psi_n \partial_z\psi_n|$ in the radiative region (see Equation
\ref{eqn:flux_expand}).  However, if one makes the assumptions explained in the Appendix, where the spatial dependence
of the eigenfunctions and their dispersion relationship become simple and continuous (see Appendix), one recovers
$F_0\propto S^{-1}$.  Thus, if one makes similar assumptions for the gravito-inertial waves, then $F_z/F_0 \propto 1$,
whereas with the more complex spatial dependence of the exact eigenfunctions and the more intricate dispersion
relationship it scales as $F_z/F_0 \propto Q(S) S^{-2}$ as seen in Figure \ref{fig:conv_waves}. Finally, both the flux
of the IGWs and the GIWs are always weaker than the total convective flux.

\section{Summary and Discussion}\label{sec:final}

A model of rotating convection originating with \citet{stevenson79} has been extended to include thermal and viscous
diffusions for any convective Rossby number in \citet{augustson19b}.  The scaling of the velocity and superadiabaticity in
terms of the colatitude, and Rossby number are outlined in Section \ref{sec:genframe}.  Asymptotically at low convective
Rossby number and without diffusion, these match the expressions given in \citet{stevenson79}, as well as the numerical
results found in the 3D simulations of \citet{kapyla05} and \citet{barker14}.

Here this rotating convection model has been employed to examine the excitation of gravito-inertial waves (GIWs) by two
different channels: one by interfacial excitation and another by Reynolds-stress excitation. First, the convection model
is applied to the interfacial wave excitation paradigm developed in \citet{press81}, where the gravity wave dynamics
there is replaced with the GIW wave dynamics computed in \citet{mathis14} and \citet{andre17}. Both mechanisms are
considered since, as seen in \citet{lecoanet15}, both sources of wave excitation play a role in simulations of gravity
wave excitation, with the dominant one being due to the volume integrated Reynolds stresses. Next, with a turbulent
convective velocity spectrum in hand, more sophisticated approaches allow for the computation of the wave energy flux in
the context of both more realistic variations in the Brunt-V\"{a}is\"{a}l\"{a} frequency as well as in a non-interfacial
paradigm that includes the Reynolds stresses throughout the convection zone. Such a step has been taken in this paper,
which builds upon the methods developed in \citet{belkacem09b}, \citet{lecoanet13}, \citet{mathis14}, where the gravity
wave and GIW excitation amplitudes and accompanying wave energy injection rate are computed by solving the wave equation
driven by a convective source term. This approach provides a general method of computing the wave flux that takes into
account the volumetric excitation of the waves and that includes the region in which they are potentially
evanescent. Specifically, to assess the influence of the convective Reynolds stresses and of rotation on the GIWs, a
wave energy flux estimate is constructed using an explicit computation of the amplitude for both the super-inertial and
sub-inertial waves.  The convection model of Paper I is then invoked as means of estimating the Reynolds stresses.

In the context of the wave energy flux, distinct parameter regimes have been found that depend upon the mode of
excitation (either interfacial pressure perturbations or convective Reynolds stresses), the convective Rossby number (or
alternatively the rotation rate), and the stiffness of the convective-radiative interface.  The visibility of these
regimes depends upon the colatitude selected, with the distinction between them being starkest at low latitudes near the
equator and vanishing at the poles due to impact of the Coriolis acceleration on the frequency range over which GIWs may
propagate. As depicted in Figure \ref{fig:wave_scaling}, interfacially-excited sub-inertial waves have a peak energy
flux near a critical convective Rossby number, but decay below it.  Interfacially-excited super-inertial waves, on the
other hand, have an increasing energy flux with increasing frequency and increasing Rossby number. As a means of
comparison, the influence of convective Reynolds stresses on the wave amplitude and their energy flux has been assessed
by directly employing the convection model of Paper I. The detailed behavior of the eigenfunctions appropriate for GIWs
and how they interact with the convective source is examined in \ref{sec:reynoldswaves}. A trend similar to that of
interfacial waves is found where there is a decline in the amplitude of the fluxes is found as the convective Rossby
number is decreased for both the sub and super-inertial waves. However, there is a large variation in the sub-inertial
wave flux for a given convective Rossby number, depending upon whether wave is in resonance or not, leading to the
series of peaks seen in Figure \ref{fig:conv_waves} where the flux relative to gravity waves in nonrotating convection
can be many orders of magnitude larger, but still below the convective flux.  The amplitude of the nonrotating
flux is computed using the same mathematical formalism as the gravito-inertial waves, but utilizing the proper
eigenmodes. The super-inertial waves have an increased flux at lower Rossby numbers reaching a peak at the transitional
Rossby number of $1/\sqrt{5}$ for the parameters chosen in Figure \ref{fig:conv_waves}.

If realized, these characteristics of GIWs may have substantial consequences for the transport and mixing of angular
momentum, chemical species, and heat in stellar and planetary interiors, in particular during the PMS of low-mass stars
or close to the base of their convective envelope, as well as consequences for the seismic observations of
them. According to the results presented in \S\ref{sec:interfacewaves} and shown in Figure \ref{fig:wave_scaling}, the
GIW energy flux due to interfacially-excited waves is likely to be reduced relative to the nonrotating case when the
wave Rossby number is not close to the critical one, meaning that any transport mechanisms associated with those waves
will be reduced as well.  Similarly, as discussed in \S\ref{sec:reynoldswaves} and shown in Figure \ref{fig:conv_waves},
the sub-inertial wave energy flux generally decreases at lower convective Rossby numbers, but reaches a peak at moderate
Rossby number. What is described here is the energy injection rate by turbulent convective motions into GIWs. However,
to get a more complete picture, the damping of GIWs resulting from their interactions with turbulent convection needs to
be studied (see \citet{samadi01} and \citet{belkacem09b}).  Nevertheless, the examination of Be star outbursts in
\citet{neiner20} have already pointed to tantilizing clues about the role of gravito-inertial waves in angular momentum
transport. There, a global description of gravito-inertial wave excitation is employed to compute the a wave spectrum
that matches the observations well (see their Figure 7).

Finally, the model examined here for the wave energy flux and excitation of GIWs has assumed a particular form of the
convective Reynolds stresses that is valid only in local domains.  However, this neglects global-scale shearing flows
seen in 3D convection simulations in spherical geometry \citep[e.g.,][]{brun11,augustson12,alvan15,emeriau17} and
theoretically predicted \citep{busse02,julien06,grooms10}, as well as neglecting the more extremal convective events
that can still occur frequently enough to influence the wave energy flux \citep[e.g.,][]{pratt17a,pratt17b}.  These
events have typically been modeled as collections of plumes for gravity waves
\citep[e.g.,][]{schmitt84,schatzman93,schatzman96a,pincon16,pincon17}, and they are tied closely to the interfacial
excitation model as such events are more likely to deform the average interface depth at least in the local region near
the plume. Therefore, the model can be further improved by considering more sophisticated models of the structure of the
flows, such as applying the models of rotating plumes considered in \citet{pedley68} or \citet{grooms10}. Thus, the
formalism developed here will be extended in future work to include the influence of rotation on those plumes as well as
a utilizing theoretical models for global-scale flows to better characterize GIW excitation and energy flux.

\section*{Acknowledgments} {The authors thank the referee for their very careful reading of the manuscript and their
  helpful and constructive comments. K.~C. Augustson, S. Mathis, and A. Astoul acknowledge support from the ERC SPIRE
  647383 grant and PLATO CNES grant at CEA/DAp-AIM.  The authors also thank Q. Andr\'{e}, U. Lee, C. Neiner,
  C. Pin\c{c}on, and V. Prat for fruitful conversations.}

\appendix{}

\section{Gravity Wave Flux in the Nonrotating and Low-Frequency Limit}

With the appropriate limit and the assumptions made in \citet{goldreich90} and \citet{lecoanet13}, which also are
similar to those made in section 4 of this paper and in \citet{press81}, one can show that they are equivalent. To do
this, recall that the definition of the flux given in Equation \ref{eqn:flux_def} is

\begin{equation}
F_z = \rho_0 \frac{f^2-\omega^2}{2\omega k_n^2} \left|A_n^2 \psi_n \partial_z \psi_n\right|.
\end{equation}

\noindent In the non-rotating limit this becomes

\begin{equation}
F_z = -\rho_0 \frac{\omega}{2k_n^2}  |A_n^2 \psi_n \partial_z \psi_n|.
\end{equation}

\noindent In the asymptotic limit of low-frequency gravity waves the JWKB approximation can be applied where
$\partial_z\psi_n \approx i k_V \psi_n$, so the flux becomes

\begin{equation}
F_z = -\rho_0 \frac{\omega k_V}{2k_n^2}  |A_n \psi_n|^2.
\end{equation}

\noindent In this limit, the vertical wavenumber is approximately

\begin{equation}
k_V = \frac{N}{\omega} k_n.
\end{equation}

\noindent So, then it can be seen that

\begin{equation}
F_z = -\rho_0 \frac{N}{2k_n}  |A_n|^2 \psi_n^2.
\end{equation}

\noindent Now, noting that $v_w = A_n \psi_n$ is the vertical velocity of the wave, one has that

\begin{equation}
F_z = -\rho_0 \frac{N}{2k_n} v_w^2.
\end{equation}

\noindent Making the assumptions of \citet{goldreich90}, \citet{lecoanet13}, and \citet{press81}, one has that $k_n
\approx k_c$, $\omega \approx \omega_c$ and $v_w \approx \omega_c^2/(k_c N)$, where the subscript $c$ indicates the wave
vector ($k_c$) and overturning frequency ($\omega_c$) of the convection.  To obtain this expression of $v_w$, we
consider the low-frequency regime where the ratio between the vertical and the horizontal components of the gravity
waves' velocity is given approximately by $v_w/v_h \approx \omega/N$. In addition, it is assumed, as in Equation (36) of
\citet{press81} and in Equation (49) of \citet{lecoanet13}, that the horizontal wave velocity is given by
$v_h \approx v_c = \omega_c/k_c$. Therefore, the previous expression becomes

\begin{equation}
F_z \approx -\rho_0 \frac{N}{2k_c} \left(\frac{\omega_c^2}{k_c N}\right)^2
     = -\rho_0 \frac{\omega_c^4}{2 k_c^3 N}.
\end{equation}

\noindent Now, since $u_c = \omega_c/k_c$ (the convective velocity), one has that

\begin{equation}
F_z \propto \rho_0 u_c^3 \frac{\omega_c}{2N} \propto F_c M,
\end{equation}

\noindent where $M = \omega_c/N = S^{-1}$ is the Mach number or the inverse stiffness ($S$) and $F_c$ is the convective
flux.  Hence, under these limits and assumptions, the flux definitions have the same scaling. Note that within these
assumptions the influence of the spatial behaviour of the eigenmodes is not apparent because
$|\psi_n|^2 \approx |e^{i k_V z}|^2 = 1$, which is not the case for the exact solutions used in Equation \ref{eqn:flux_expand}.

\bibliography{waves}

\end{document}